\documentclass{jfm}

\usepackage{graphicx}
\usepackage{epstopdf,epsfig}
\usepackage{newtxtext}
\usepackage{newtxmath}
\usepackage{natbib}
\usepackage{hyperref}
\usepackage[amssymb]{SIunits}
\usepackage{upgreek}

\usepackage{placeins}

\usepackage{color}

\hypersetup{
    colorlinks = true,
    urlcolor   = blue,
    linkcolor  = magenta,
    citecolor  = blue,
}

\newcommand{\RomanNumeralCaps}[1]

\title{Buoyancy-driven flow regimes for a melting vertical ice cylinder in saline water}

\author{Dehao Xu\aff{1}\footnotemark[1],
  Simen T. Bootsma\aff{1}\footnotemark[1],
  Roberto Verzicco\aff{1,2,3},
 Detlef Lohse\aff{1,4} \footnotemark[2]
 \and Sander G. Huisman\aff{1} \footnotemark[2]\footnotetext[1]{The two authors contributed equally to this paper.} \footnotetext[2]{Email addresses for correspondence: d.lohse@utwente.nl, s.g.huisman@utwente.nl}}

\affiliation{\aff{1}Physics of Fluids Department, Max Planck Center for Complex Fluid Dynamics, and J. M. Burgers Centre for Fluid Dynamics, University of Twente, P.O. Box 217, 7500AE Enschede, The Netherlands
\aff{2}Dipartimento di Ingegneria Industriale, University of Rome `Tor Vergata', Roma 00133, Italy
\aff{3}Gran Sasso Science Institute - Viale F. Crispi, 7 67100 L'Aquila, Italy
\aff{4}Max Planck Institute for Dynamics and Self-Organization, Am Fa\ss berg 17, 37077 G{\"o}ttingen, Germany}

\begin{document}
\maketitle

\begin{abstract}
The presence of salt in seawater significantly affects the melt rate and morphological evolution of ice. This study investigates the melting process of a vertical cylinder in saline water using a combination of laboratory experiments and direct numerical simulations. The two-dimensional (2D) direct numerical simulations and 3D experiments achieve thermal Rayleigh numbers up to $\text{Ra}_{T}= \mathcal{O}\left(10^{9}\right)$ and saline Rayleigh numbers up to $\text{Ra}_{S}=\mathcal{O}\left(10^{12}\right)$. Some 3D simulations of the vertical ice cylinder are conducted at $\text{Ra}_{T}= \mathcal{O}\left(10^{5}\right)$ to confirm that the results in 2D simulations are qualitatively similar to those in 3D simulations. The mean melt rate exhibits a non-monotonic relationship with ambient salinity. With increasing salinity, the mean melt rate initially decreases towards the point where thermal and saline effects balance, after which it increases again. Based on the ambient salinity, the flow can be categorized into three regimes: temperature-driven flow, salinity-driven flow, and thermal-saline competing flow. In the temperature-driven and competing flow regimes, we find that the mean melt rate follows a $\text{Ra}_{T_d}^{1/4}$ scaling, where the subscript $d$ denotes a response parameter. In contrast, in the salinity-driven flow regime, we see a transition from a $\text{Ra}_{T_d}^{1/4}$ to a $\text{Ra}_{T_d}^{1/3}$ scaling. Additionally, the mean melt rate follows a $\text{Ra}_{S_d}^{1/3}$ scaling in this regime. The ice cylinder develops distinct morphologies in different flow regimes. In the thermal-saline competing flow regime, distinctive scallop (dimpled) patterns emerge along the ice cylinder due to the competition between thermal buoyancy and saline buoyancy. We observe these scallop patterns to migrate downwards over time, due to local differences in the melt rate, for which we provide a qualitative explanation.
\end{abstract}

\begin{keywords}
\end{keywords}


\newpage
\section{Introduction}\label{sec:introduction}
The process of melting is important across various fields in geophysics and industry, encompassing scenarios such as iceberg and ice shelf melting \citep{Huppert1978,Epsterin1983,Dutrieux2014,Ristroph2018}, food industry \citep{Rahman2020} and phase-change dynamics of materials \citep{Dhaidan2015}. In particular, with the persistence of global warming, a solid understanding of the melting dynamics of glaciers and sea ice becomes increasingly significant. Despite this urgency, predictions by present climate models consistently underestimate the observed decline in Arctic sea ice \citep{Stroeve2007}, and Alaskan glaciers are melting at rates surpassing predictions by orders of magnitude \citep{Sutherland2019}. Hence, refining existing climate models is imperative to improve their accuracy in projecting the observed loss of glaciers and sea ice. This requires a deeper understanding of ice sheet-ocean interactions at a fundamental level, given that the governing physical mechanisms underlying these interactions remain inadequately understood \citep{Truffer2016, Malyarenko2020, Cenedese2023, Rosevear2025}.

Melting of ice is a Stefan problem \citep{Rubinstein1971}, in which the evolution of the ice-water interface is obtained by solving heat equations in both the liquid phase and the solid phase. Since the interface changes over time, melting is considered a free boundary problem, as opposed to fixed boundary problems where the liquid-solid interface remains stationary. During melting of ice in water, the density of ambient water near the ice constantly changes due to temperature gradients induced by the melting ice. Depending on the ice geometry and the ambient conditions of the water, such density variations will result in either a stable or an unstable stratification. The latter leads to a convective flow, which in turn affects the heat flux to the ice and, consequently, the evolution of the ice-water interface. Therefore, the convective flow and the interface evolution are two-way coupled, significantly complicating the problem. While buoyancy-driven convective flows have been investigated extensively in fixed boundary problems \citep{bejan1993,Kadanoff2001,Ahlers2009,Lohse2010,Schlichting2017,Xia2023,Loshe2023,Lohse2024}, they have been addressed less frequently in evolving boundary problems, such as the melting of ice.

Out of the limited amount of studies under controlled conditions from a fundamental point of view, most considered a single-component ambient liquid, in which temperature gradients are the only source of buoyancy \citep{Favier2019,Wang2021,Weady2022,Yang2023}. Such temperature-driven flows are influenced by several factors, such as the presence of a forced flow \citep{Bushuk2019, Couston2021, Hester2021} and rotation of the system \citep{Ravichandran2021,Toppaladoddi2021,Ravichandran2022}. In particular, the density anomaly of water causes rich flow phenomena \citep{Wang2021}, including the formation of recirculating eddies that `sculpt' wave-like patterns on the ice surface \citep{Weady2022}.

Such wave-like or dimpled patterns that form on an ablative solid are generally referred to as scallops. These features are common in various landscapes, having been observed in limestone caves, granular beds, solidified sugar, sea ice, river ice, and the undersides of icebergs (e.g. \citet{Curl1966, Blumberg1974, Thomas1979, Wykes2018, Hobson2011}). Laboratory experiments on a horizontal ice layer, subjected to turbulent forced convection, have shown that scallops significantly enhance heat transfer to ice in freshwater \citep{Gilpin1980}. Unlike the scallops resulting from flow instabilities as reported by \citet{Weady2022}, those in forced convection arise from the positive feedback between shear production and ice surface geometry \citep{Claudin2017, Bushuk2019}. An extensive overview of the work done in this area is given by the recent review of \citet{Du2024}. They discuss the interaction between flow and ice morphology in a large number of configurations, among which a melting horizontal ice layer above saline water and a melting vertical ice wall in water with a vertical salinity gradient. Yet, the morphology of a vertical ice face melting in homogeneous saline water, as considered here, remained elusive. \citet{Du2024} also point out that although field measurements are essential to capture the full complexity of such phenomena, laboratory experiments and fully resolved simulations offer a controlled way to understand the small-scale physical processes. Further research on scallops is needed to understand their impact on the melt rate of icebergs, sea ice, and glacier ice, particularly concerning their effect on heat transport to the ice and the mechanisms leading to their formation.

Compared to melting in freshwater, the melting dynamics of ice in seawater are significantly complicated by the interplay of temperature and salinity, upon which both the density of water and the melting point of ice depend. The process is further complicated by the significantly slower diffusion of salt in water compared to diffusion of heat. Specifically, salt diffuses approximately one hundred times more slowly than heat (Lewis number $\text{Le} = \kappa_T/\kappa_S = \mathcal{O}(100)$), resulting in a saline boundary layer that is ten times thinner than the thermal boundary layer \citep{Schlichting2017}. For typical conditions and material properties of water, we therefore have the situation of a triply-nested boundary layer. The saline boundary layer is nested in the thermal boundary, which is nested in the kinetic boundary layer. These differences in boundary layer thicknesses give rise to various flow phenomena, collectively referred to as double-diffusive convection (DDC) \citep{Radko2013}.

In the context of melting, DDC induces complex flow patterns that significantly affect the melt rate and morphological dynamics of ice. For a homogeneous ambient salinity, DDC permits the existence of several flow regimes, contingent upon the ambient temperature and salinity \citep{Josberger1981, Carey1982, Sammakia1983}. Specifically, the flow adjacent to a melting vertical ice wall has been observed to be fully upward at low temperatures and high salinities, fully downward at high temperatures and low salinities, and bi-directional at intermediate temperatures and salinities \citep{Carey1982}. For low ambient temperatures ($T_\infty < $ \unit{6}{\celsius}) and ocean salinity, \citet{Kerr2015} have reported a theoretical prediction for the dissolution velocity of a vertical ice wall that is in good agreement with their experiments on a one-meter-tall ice wall. They conclude that the dissolution velocity $\mathscr{V}$ should scale with the driving temperature difference $\Delta T_d$ as $\mathscr{V} \propto \Delta T_d^{4/3}$. In the presence of a salinity gradient, DDC induces layering of the flow adjacent to a vertical ice block. The resultant layers contain convection cells that etch a regular pattern in the ice \citep{Huppert1980,Sweetman2024}.

While such experiments have been conducted since the 1980s, numerical simulations of melting ice in saline water have been comparatively limited. Due to the low diffusivity of salt, simulations require a high resolution to accurately resolve the flow, leading to a high computational cost. Therefore, most numerical simulations of melting ice in saline water have only been performed in recent years. One study that performed direct numerical simulations in water at ocean salinity and temperatures below \unit{6}{\celsius}, showed the complexity of turbulent dissolution of a vertical ice face \citep{Gayen2016}. Their results show good agreement with the $\mathscr{V} \propto \Delta T_d^{4/3}$ reported by \citet{Kerr2015}. Simulations of ice in a salinity gradient have revealed a similar layering of the flow as observed in the early experiments by \citet{Huppert1980}, as well as a non-monotonic dependence of the melt rate on ambient salinity \citep{Yang2023b, Wilson2023}.. 

Despite the notable advancements in previous studies, several pressing questions remain unanswered. First, the development of theoretical models or scaling laws for the mean melt rate is essential to improve predictions of melting dynamics. Second, there is a lack of understanding regarding the physical mechanisms that lead to the observed morphology evolution during the melting of ice in saline water. Hence, in this study, we explore the melting process of a vertical ice cylinder in saline water through a combination of numerical simulations and laboratory experiments. Our investigation primarily focuses on establishing scaling laws for the mean melt rate, and describing the interplay between the convective flows and the morphology dynamics.

The paper is organized as follows. In § \ref{sec:equations}, we introduce the underlying equations and the control and response parameters. In § \ref{sec:methods}, we describe the experimental and numerical methodologies. In § \ref{sec:melt_rate}, we discuss the scaling of the mean melt rate with both thermal and saline Rayleigh numbers. In § \ref{sec:flow_regimes}, we report on the influence of ambient salinity on the presence of different flow regimes and their effect on the ice morphology. After that, in § \ref{sec:morphology}, we elaborate on the properties of the ice scallops. Finally, we provide conclusions and an outlook in § \ref{sec:conclusion}.

\section{Underlying equations and the control and response parameters}\label{sec:equations}
In the present study, we assume the flow under consideration to be incompressible, such that the density field $\rho $ is determined by the temperature ($T\left( \boldsymbol{x},t \right)$) and salinity ($S\left( \boldsymbol{x},t \right)$) fields. The governing equations for our direct numerical simulation, incorporating the phase-field variable $\phi$ needed for melting, along with the incompressibility condition $\boldsymbol{\nabla }\cdot \boldsymbol{u}=0$, read as follows:
\begin{align}
    \frac{\partial \boldsymbol{u}}{\partial t}+\boldsymbol{u}\cdot \boldsymbol{\nabla }\boldsymbol{u} &= -\frac{\boldsymbol{\nabla }p}{\rho _{0}}+\nu \boldsymbol{\nabla }^{2}\boldsymbol{u}-\frac{g{\rho }'}{\rho _{0}} \boldsymbol{e}_{y}-\frac{\phi \boldsymbol{u}}{\eta }, \\
    \frac{\partial T}{\partial t}+\boldsymbol{u}\cdot \boldsymbol{\nabla }T &= \kappa _{T}\boldsymbol{\nabla }^{2}T+\frac{\mathcal{L}}{c_{p}}\frac{\partial \phi }{\partial t}, \\
    \frac{\partial S}{\partial t}+\boldsymbol{u}\cdot \boldsymbol{\nabla }S &= \kappa _{S}\boldsymbol{\nabla }^{2}S+\frac{1}{1-\phi +\delta }\left ( S\frac{\partial \phi }{\partial t}- \kappa _{S}\boldsymbol{\nabla }\phi \cdot \boldsymbol{\nabla }S\right ), \label{eq:saleqn}\\
    \frac{\partial \phi }{\partial t} &= C\boldsymbol{\nabla }^{2}\phi -\frac{C}{\varepsilon ^{2}}\phi \left ( 1-\phi  \right )\left ( 1-2\phi +\frac{\varepsilon }{\gamma  }\left ( T+mS \right )\right ).
    \label{eqn: phieqn}
\end{align}
Here, $\boldsymbol{u}\left( \boldsymbol{x},t \right)$ is the velocity field, ${\rho }'=\rho -\rho _{0}$ is the fluctuating density from a reference value $\rho _{0}$, $p$ is the kinematic pressure, $\nu $ is the kinematic viscosity of the liquid, $g$ is the gravitational acceleration in the vertical direction $\boldsymbol{e}_{y}$, $\kappa _{T}$ and $\kappa _{S}$ are the diffusivities of temperature and salinity, respectively. In simulations, we employ the phase-field method \citep{Hester2020} to simulate the melting process of a solid in a multi-component fluid, which has been extensively utilized in numerous prior numerical investigations of melting ice \citep{Favier2019,Hester2021,Couston2021,Yang2023b,Yang2023}. In this method, the phase field variable $\phi $ is integrated in time and space, and smoothly transitions from a value of 1 in the solid to a value of 0 in the liquid, with the interface located at $\phi =1/2$. $\mathcal{L}$ is the latent heat of fusion, and $c_{p}$ is the speciﬁc heat capacity. The liquidus slope $m=\unit{0.056}{\kelvin/(\gram\per\kilo\gram)}$ reflects how the presence of salinity at the ice-water interface lowers the local melting temperature. 

The phase-field model employed in this study follows the second-order formulation of \citet{Hester2020} and properly reflects the Gibbs–Thomson effect \citep{Hester2020}. The parameters $\varepsilon $, $\eta $, $C$, $\gamma $ and $\delta $ all refer to the phase-field model, which can be explained as follows: The parameter $\varepsilon $ is used to measure the diffuse interface thickness, which is set to be equal to the grid spacing following the convergence test in \citet{Favier2019}. The limit $\varepsilon \rightarrow 0$ leads to the exact Stefan boundary conditions for $S$ and $T$ at the liquid-solid interface:
\begin{align}
   S^{\left ( l \right )}u_{n} &= - \kappa_{S} \frac{\partial S^{\left( l \right)}}{\partial n} \label{eqn: stS} \\
    \mathcal{L}u_{n} &= c_{p}\kappa _{T}\left ( \frac{\partial T^{\left( s \right)}}{\partial n} - \frac{\partial T^{\left ( l \right )}}{\partial n} \right ),
    \label{eqn: stT} 
\end{align}
where $u_{n}$ is the normal velocity of the interface between the solid and the liquid phases, $n$ represents the normal direction of the interface, and the superscripts $\left(s\right)$ and $\left(l\right)$ represent the solid and liquid phases, respectively. The penalty parameter $\eta $ is used to decay the velocity to zero in the solid phase and its value is set to be equal to the time interval \citep{Hester2020}. Furthermore, a direct forcing method is applied to set the velocity to zero for $\phi > 0.9$ in order to avoid spurious motions in the solid phase \citep{Howland2022}. The diffusivity $C$ of the phase ﬁeld equation (\ref{eqn: phieqn}) is deﬁned by $C=6\gamma \kappa _{T}/\left (5\varepsilon \mathcal{L}  \right )$, where $\gamma $ is the surface energy coefficient related to the Gibbs--Thompson effect \citep{Hester2020,Howland2022,Yang2023b}. We set $C=1.2\kappa_T$ and $\varepsilon \Delta T/\gamma =10$, where $\Delta T$ is the initial temperature difference between the solid and liquid phases. The small parameter $\delta \ll 1$ is solely used to stabilise the terms on the right hand side of the salinity equation (\ref{eq:saleqn}) \citep{Howland2022}. Further description and validation of the phase-field method can be found in \citet{Hester2020} and \citet{Howland2022}.

In the simulations, the initial temperature of the water surrounding the ice is set to $T_{\infty }=\unit{20}{\celsius}$. Since this is much larger than the maximum-density temperature $T_c=\unit{4}{\celsius}$, the Oberbeck--Boussinesq approximation is employed, ignoring the density anomaly that occurs at \unit{4}{\celsius} for water. The fluctuating density is assumed to depend linearly on two scalar fields, namely the temperature $T$ and salinity $S$,
\begin{align}
    \rho'=\rho_0 \left( \beta_{S}S - \beta_{T}T \right),
    \label{eqn:rhop}
\end{align}
where $\beta_T $ is the thermal expansion coefficient, and $\beta_S $ is the haline contraction coefficient. We verified that the results of simulations considering the density anomaly are nearly identical to those of simulations ignoring the density anomaly. Obviously, this does no longer hold for an ambient water temperature $T_{\infty } \approx \unit{4}{\celsius}$.

The system can be controlled using five dimensionless control parameters, namely the Rayleigh numbers of temperature and salinity, the Prandtl number, the Schmidt number, and the Stefan number:
\begin{align}
    \text{Ra}_{T}=\frac{g\beta _{T} \Delta TH^{3}}{\nu \kappa _{T}},\,\text{Ra}_{S}=\frac{g\beta _{S}\Delta SH^{3}}{\nu \kappa _{S}},\,\text{Pr}=\frac{\nu }{\kappa _{T}},\,\text{Sc}=\frac{\nu }{\kappa _{S}},\,\text{Ste}=\frac{c_{p}\Delta T }{\mathcal{L}},
    \label{eqn:control_params}
\end{align}
where $\Delta T=T_{\infty }-T_{s}$ and $\Delta S=S_{\infty }-S_{s}$ are the temperature and salinity differences between the surrounding water and the ice, respectively. Here $T_s=\unit{0}{\celsius}$ is the temperature of the ice, which is the equilibrium melting temperature for freshwater. Furthermore, $S_{\infty }$ and $S_s=\unit{0}{\gram\per\kilo\gram}$ are the salinity of the ambient water and the ice, respectively.

The Lewis number $\text{Le}$ and the density ratio $R_\rho$ corresponding to equations (\ref{eqn:control_params}) are given by:

\begin{equation}
    \text{Le}=\frac{\kappa _{T}}{\kappa _{S}}=\frac{\text{Sc}}{\text{Pr}}, \quad R_\rho \equiv \frac{\text{Ra}_S \text{Pr}}{\text{Ra}_T \text{Sc}} = \frac{\beta_S \Delta S}{\beta_T \Delta T},
    \label{eqn:Le_Pr_Rp}
\end{equation}
Here the density ratio $R_\rho$ is introduced to quantify the ratio between the strengths of the saline and thermal buoyancies.

While the experiments and simulations are initialized using the control parameters $\mathrm{Ra}_T$ and $\mathrm{Ra}_S$, the system is physically driven by the differences in temperature and salinity between the ambient water and the ice-water interface, which are $\Delta T_d = T_\infty - T_i$ and $\Delta S_d = S_\infty - S_i$, respectively. Here $T_{i}$ and $S_{i}$ denote the temperature and salinity at the ice–water interface. Accordingly, the corresponding Rayleigh numbers are given by:

\begin{align}
    \text{Ra}_{T_d}=\frac{g\beta _{T} \Delta T_{d}H^{3}}{\nu \kappa _{T}},\,\text{Ra}_{S_d}=\frac{g\beta _{S}\Delta S_{d}H^{3}}{\nu \kappa _{S}}.
    \label{eqn:response_params}
\end{align}

Since $T_i$ and $S_i$ are actively changing during the experiments and simulations, $\mathrm{Ra}_{T_d}$ and $\mathrm{Ra}_{S_d}$ are response parameters, as opposed to the control parameters $\mathrm{Ra}_T$ and $\mathrm{Ra_S}$.

For the experiments both $T_i$ and $S_i$ are assumed to be zero, which is a reasonable assumption for the relatively high ambient temperature considered in this study ($T_\infty \approx$ \unit{20}{\celsius}). In this case, \citet{Josberger1981} experimentally verified that the interface salinity is less than 1\% of the ambient salinity at least up to $S_\infty = $ \unit{35}{\gram\per\kilo\gram}. According to the liquidus condition, the interface temperature is therefore less than 0.1\% of the ambient temperature. Note that the interface salinity becomes more relevant as the ambient temperature approaches the freezing temperature \citep{Kerr2015}.

For the simulations, the interface temperature $T_i$ and salinity $S_i$ are computed as time-averaged surface-integral values over the ice–water interface. Specifically, $S_i$ is computed by $ S_i=\left< \int \left |\nabla \phi \right |Sd\Omega /\int \left |\nabla \phi \right |d\Omega \right>_t$, where $\Omega $ represents the computational domain. The interface temperature $T_i$ is subsequently determined from $S_i$ via the liquidus condition $T_i=T_{m}-mS_{i}$, where $T_{m}=\unit{0}{\celsius}$ is the melting point of fresh water.

In this paper, the normalised mean melt rate $\widetilde{f}$ is defined by $\widetilde{f}=t_{D}/t_{m}$, where $t_{m}$ represents the time needed to melt $V_{m}=70\%$ of the initial volume in 3D simulations and experiments and of the initial area in 2D simulations. Here $t_{D}=H^{2}/\kappa _{T}$ is the diffusion time scale of temperature. We also tried to use different values of $V_{m}$ to calculate the melt time $t_{m}$ and the normalised mean melt rate $\widetilde{f}$ in 2D simulations. This only resulted in a change of the absolute values of $t_{m}$ and $\widetilde{f}$, but the trend and scaling remained the same. This invariance is further corroborated by the approximately constant instantaneous melt rate observed in the 2D simulations, as discussed in Appendix \ref{appC}. Therefore, the value of $V_{m}$ has no impact on the qualitative results of the present study.

According to the Stefan boundary condition for salinity (equation (\ref{eqn: stS})), the dimensionless instantaneous melt speed $\widetilde{u}_{n}$ can be related to the dimensionless instantaneous salt flux ${\text{Nu}}_{S}$:

\begin{equation}
    \widetilde{u}_{n}=\frac{u_{n}}{U_{0}}=-\frac{1}{U_{0}}\frac{\kappa _{S}}{S}\frac{\partial S}{\partial n}=\frac{{\text{Nu}}_{S}}{\text{Le}}\frac{S_{\infty }-S_{i}}{S_{i}}\frac{H}{\delta _S(t)},
    \label{eqn: NuS}
\end{equation}
where $U_{0}=H/t_{D}$ is the thermal diffusion velocity scale, $\delta _S\left ( t \right )$ is the saline boundary layer thickness at time $t$, and ${\text{Nu}}_{S}=-\left ( \partial S/\partial n \right )/\left ( \Delta S_{d}/\delta _S(t) \right )$. Since the mean melt rate $\widetilde{f}$ is per definition equal to the time-average of $\widetilde{u}_{n}$, equation (\ref{eqn: NuS}) expresses that $\widetilde{f}$ must be proportional to the dimensionless time-averaged salt flux $\overline{\text{Nu}}_{S}$. 

Similar to equation (\ref{eqn: NuS}), we relate the dimensionless instantaneous melt speed $\widetilde{u}_{n}$ to the dimensionless instantaneous heat flux $\text{Nu}_{T}$, according to the Stefan boundary condition for temperature (equation (\ref{eqn: stT})):

\begin{align}
    \widetilde{u}_{n}=\frac{u_{n}}{U_{0}}=-\frac{1}{U_{0}}\frac{\kappa_{T} C_{p}}{\mathcal{L}}\frac{\partial T}{\partial n}=\text{Ste}\,\text{Nu}_{T} \frac{T_{\infty }-T_{i}}{T_{\infty }} \frac{H}{\delta _T(t)},
    \label{eqn: NuT}
\end{align}
where ${\text{Nu}}_{T}=-\left ( \partial T/\partial n \right )/\left ( \Delta T_{d}/\delta _T(t) \right )$, and $\delta _T\left ( t \right )$ is the thermal boundary layer thickness at time $t$. Throughout the entire melting process, the ice temperature $T_{\mathrm{ice}}$ remains within the range of $-0.4\,^{\circ}\mathrm{C}$ to $0\,^{\circ}\mathrm{C}$, indicating that the heat flux from the interior of the ice to the ice-water interface is much smaller than the heat flux from the surrounding water to the interface. Therefore, we assume that the term $\partial T^{\left( s \right)}/\partial n$ is much smaller compared with the term $\partial T^{\left( l \right)}/\partial n$. Accordingly, the mean melt rate $\widetilde{f}$ should be proportional to the dimensionless time-averaged heat flux $\overline{\text{Nu}}_{T}$. Several previous studies \citep{Josberger1981,Kerr1994,Kerr2015,Howland2023} have reported that the ratio of thermal to solutal boundary layer thicknesses scales as $\delta_T(t)/\delta_S(t) \propto \mathrm{Le}^{1/2}$ for Lewis numbers of order $\mathrm{Le} = \mathcal{O}(100)$. Adopting this scaling, we assume $\delta_T(t)/\delta_S(t) = \mathrm{Le}^{1/2}$, and we find the relation between $\overline{\text{Nu}}_{S}$ and $\overline{\text{Nu}}_{T}$, namely \citep{Holland1999,Couston2024}
\begin{equation}
\overline{\text{Nu}}_{S}=\text{Ste}\,\text{Le}^{1/2}\,\frac{S_{i}}{S_{\infty }-S_{i}}\frac{T_{\infty }-T_{i}}{T_{\infty }}\,\overline{\text{Nu}}_{T}.
    \label{eqn: NuSNuT}
\end{equation}

\begin{figure}\centering
         \includegraphics[width=\linewidth]{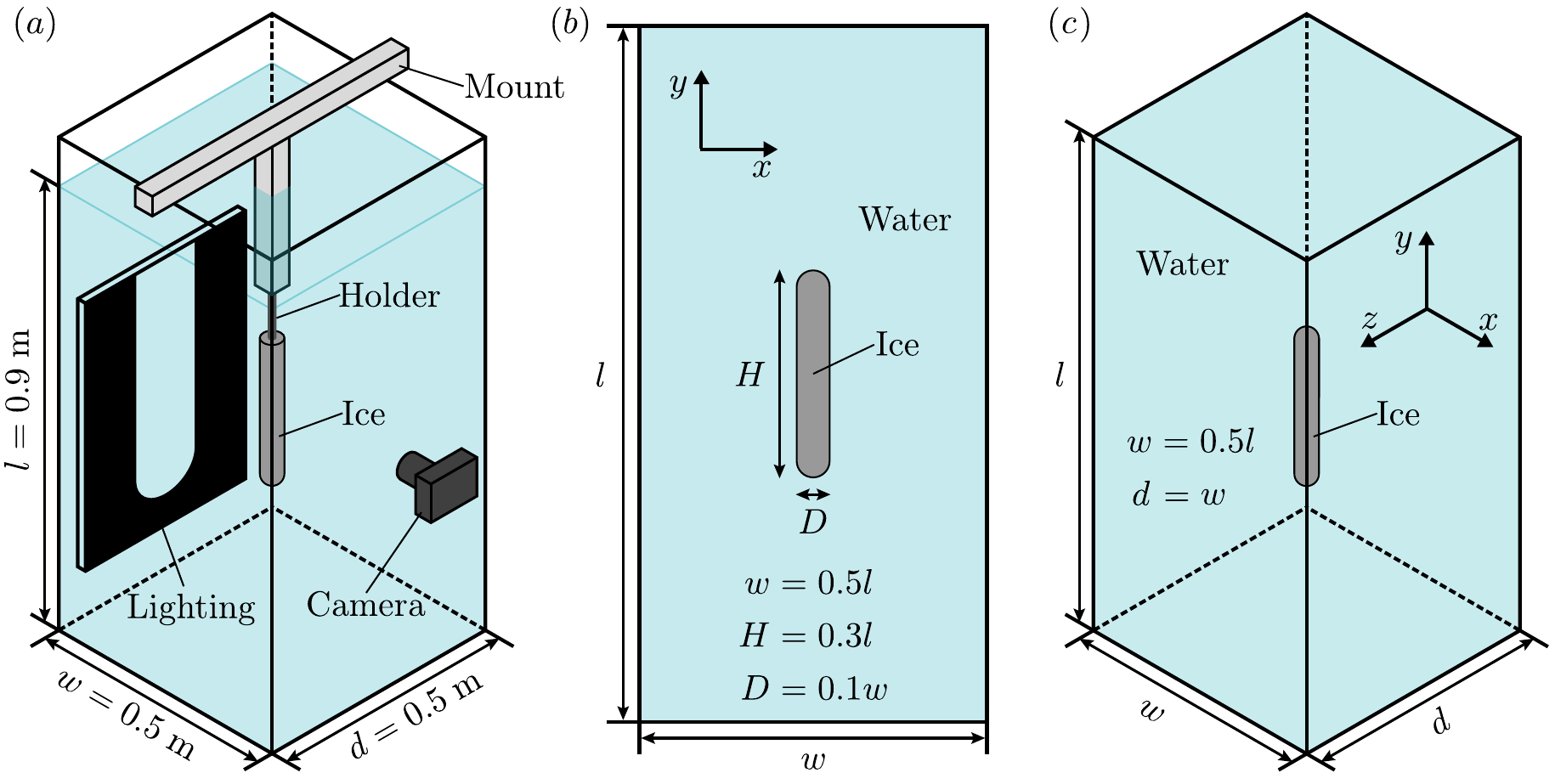}
	\caption{(\textit{a}) Schematic of the experimental setup. A glass tank is filled with water. The cylinder is placed at the center of the tank and is kept in place using a holder made from POM which is thermally poorly conducting. The ice has a hemispherical cap at the bottom, and is flat at the top. (\textit{b}) and (\textit{c}): Schematic of the numerical setup in (\textit{b}) 2D and (\textit{c}) 3D simulations. The object is fixed at the center of the domain. The ice has a circular cap at both ends in the 2D simulations, and hemispherical caps for the 3D cases.}
	\label{fig:fig1}
\end{figure}

\section{Experimental and numerical methods}\label{sec:methods}

\subsection{Experimental method and setup}

The experimental setup consisted of a $l \times w \times d = \unit{90}{\centi\meter} \times \unit{50}{\centi\meter} \times \unit{50}{\centi\meter}$ glass water tank, as sketched in figure \ref{fig:fig1}(\textit{a}). The tank was filled with water and NaCl was dissolved to increase the salinity. Ambient water temperature and salinity were measured using a StarOddi DST CT probe with a temperature accuracy of $\unit{\pm 0.1}{\kelvin}$ and a salinity accuracy of $\unit{\pm 1}{\gram\per\kilo\gram}$. The difference in composition between seawater and our NaCl solution is neglected in the salinity measurement, which adds an uncertainty of at most $\unit{0.1}{\gram\per\kilo\gram}$ \citep{teos10primer}. During all experiments, the ambient water temperature and salinity were measured to be within \unit{0.1}{\kelvin} and \unit{0.1}{\gram\per\kilo\gram} from the reported temperature and salinity, respectively. Prior to each experiment, we thoroughly mixed the water using a stirrer to achieve uniform temperature and salinity throughout the tank, then waited for the flow to quiesce. We used a \unit{30}{\centi\meter} long ice cylinder, rounded on the bottom side, with a diameter of \unit{5}{\centi\meter} for all experiments. Before each experiment, the ice was taken out of the freezer and left to equilibrate to room temperature for approximately 30 minutes, after which the ice core temperature was measured to be $T_s = \unit{0.0}{\celsius} \pm \unit{0.5}{\kelvin}$. The ambient conditions for all experiments are listed in table \ref{tab: tab1}. Here, we used equation (\ref{eqn:control_params}) to compute $\text{Ra}_T$ and $\text{Ra}_S$, for which the density $\rho$, kinematic viscosity $\nu$, and thermal diffusivity $\kappa_T$ were obtained from the empirical correlations of \citet{Sharqawy2010}, evaluated at ambient temperature and salinity. Using the same expression for $\rho$, the thermal expansion coefficient $\beta_T = - (1/\rho)\partial\rho/\partial T$ and the haline contraction coefficient $\beta_S = (1/\rho)\partial\rho/\partial S$ were obtained, and evaluated at ambient temperature and salinity.

\begin{figure}\centering
    \includegraphics[width=0.99\linewidth]{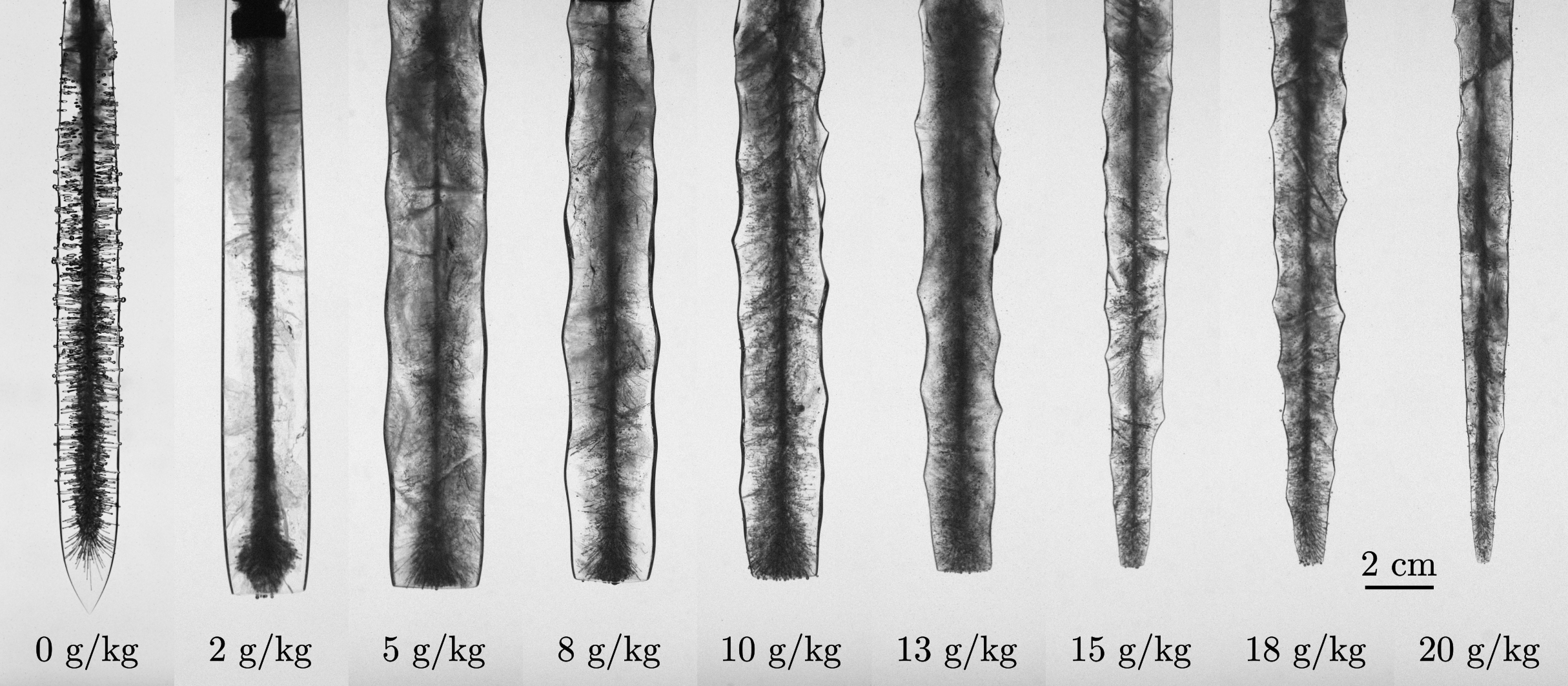}
\caption{Snapshots of nine experiments after \unit{17}{\minute} of melting, with increasing ambient salinity from left to right. Black regions inside the ice are pockets of entrapped air.}
\label{fig:fig2}
\end{figure}

The shape evolution of the ice cylinder was quantified through boundary tracking, i.e., extraction of the ice-water interface from photographs taken at a regular time interval. This was done using a single camera pointing at the side of the cylinder, so only the 2D projection of the cylinder was obtained. We used a Nikon D850 camera with a Sigma \unit{105}{\milli\meter} macro objective, resulting in a resolution of \unit{42}{\mu\meter\per px}. We captured photographs at a \unit{10}{\second} interval for the experiments up to $\Delta S = \unit{20}{\gram\per\kilo\gram}$ and a \unit{5}{\second} interval for the experiments with $\Delta S = \unit{34}{\gram\per\kilo\gram}$ and $\Delta S = \unit{81}{\gram\per\kilo\gram}$, because of the higher melt rate. As illustrated in figure \ref{fig:fig1}(\textit{a}), the cylinder was lit from behind using a computer screen, on which the approximate shape of the cylinder was displayed. The width, height, and curvature of this shape were manually adjusted during the experiment to ensure sufficient contrast along the edge of the cylinder for the entire duration of the experiment. Examples of photographs taken during the experiments are shown in figure \ref{fig:fig2}. Each photograph was desaturated and then binarised using a global threshold, after which the edge contours were extracted using the border following algorithm developed by \cite{Suzuki1985}. Using these contours, the cylinder volume was computed as a vertical sum of disks, each with a height of 0.5 mm and a diameter based on the contours. The black regions in figure \ref{fig:fig2} show the presence of air bubbles inside the ice. We expect the effect of this entrapped air on our results to be minimal. By measuring the ice density, we estimated the volume fraction of air to be 1\%. Furthermore, most of the air is located in the center of the cylinder, as the ice was frozen radially inward. In our analysis we only use the 70\% of the volume that melts first, so we expect both the mean melt rate and the emergence of patterns to be unaffected by the entrapped air.\\

\begin{table}
	\begin{center}
		\def~{\hphantom{0}}
\setlength{\tabcolsep}{6mm}
		\begin{tabular}{ccccc}
			$\unit{S_\infty}{(\gram\per\kilo\gram)}$ & $\unit{T_\infty}{(\degreecelsius)}$ & $\text{Ra}_{S}$ & $\text{Ra}_{T}$ & $R_\rho$\\ [4pt]
                \hline
                $0.0$ &$17.8$ & $0.0$ & $5.7\times 10^{9}$ &  $0.0$ \\[4pt]
                $2.4$ &$15.9$ &$3.0\times 10^{11}$ & $4.4\times 10^{9}$ &   $0.69$ \\[4pt]
                $4.8$ &$15.5$ & $6.0\times 10^{11}$ &$4.2\times 10^{9}$ &   $1.4$ \\[4pt]
                $7.5$ &$16.8$ & $9.5\times 10^{11}$ & $5.2\times 10^{9}$ &  $1.8$ \\[4pt]
                $9.9$ &$16.5$ & $1.2\times 10^{12}$ & $5.1\times 10^{9}$ &  $2.4$ \\[4pt]
                $12.5$ &$17.1$ &$1.6\times 10^{12}$ &$5.7\times 10^{9}$ &   $2.8$ \\[4pt]
                $15.0$ & $16.8$ & $1.9\times 10^{12}$ & $5.5\times 10^{9}$ & $3.4$ \\[4pt]
                $18.1$ &$17.4$ & $2.3\times 10^{12}$ &$6.1\times 10^{9}$ &   $3.7$ \\[4pt]
                $20.4$ &$17.1$ & $2.5\times 10^{12}$ &$6.0\times 10^{9}$ &  $4.4$ \\[4pt]
                $34.3$ &$19.4$ & $4.3\times 10^{12}$ & $8.3\times 10^{9}$ &  $5.2$\\[4pt]
                $80.8$ &$18.7$ & $9.1\times 10^{12}$ & $8.6\times 10^{9}$ &  $10.6$\\[4pt]
		\end{tabular}
		\caption{Experimental conditions of the ambient water, including the ambient salinity $S_\infty$ and ambient temperature $T_\infty$}, and their corresponding Rayleigh numbers $\text{Ra}_S$ and $\text{Ra}_T$, respectively. The last column also contains the density ratio $R_\rho$. The dimensions of the cylinder were kept constant at $H = \unit{30}{\centi\meter}$ and $D = \unit{5}{\centi\meter}$.
		\label{tab: tab1}
	\end{center}
\end{table}

\subsection{Numerical method and setup}
Both two-dimensional (2D) and three-dimensional (3D) simulations are conducted in this study, with the ice represented as a cylinder in the 3D simulations. For 2D simulations, the flow is confined to a box of height $l$ and width $w$, with the aspect ratio $\Gamma = w/l = 0.5$. For 3D simulations, the depth $d$ is the same as the width $w$. No-slip, no-heat-flux, and no-salt-flux boundary conditions are applied on all walls. A vertical ice cylinder with a height $H = 0.3l$ and diameter $D = 0.1w$ is fixed at the center of the box and surrounded by uniformly warm and salty water, as shown in figure \ref{fig:fig1}(\textit{b}, \textit{c}). The aspect ratios of the vertical ice cylinder and the box in simulations are consistent with those in experiments.

Simulations are performed using the second-order staggered finite difference code AFiD, which has been extensively validated and used to investigate a wide range of convection problems \citep{Verzicco1996,vanderPoel2015,Ostilla2015,Yang2016}. The extension of the AFiD code to study phase-change problems with the phase-field method was also validated and discussed in recent studies \citep{Liu2021,Yang2023b,Yang2023}. 

In this study, several control parameters are fixed to restrict the extensive parameter space. Specifically, the Prandtl number and Schmidt number are set at $\text{Pr}=1$ and $\text{Sc}=100$ respectively, resulting in a Lewis number $\text{Le}=100$. It is important to note that the Prandtl and Schmidt numbers ($\text{Pr}=1$ and $\text{Sc}=100$) used in the simulations are lower than those in the experiments, where $\text{Pr}\approx7$ and $\text{Sc}\approx700$. At higher Prandtl and Schmidt numbers, the thermal and saline boundary layers become significantly thinner, requiring substantially finer grid resolutions to accurately capture the boundary layer dynamics. Moreover, the mean melt rate decreases with increasing Prandtl and Schmidt numbers, leading to a longer melt time for melting a given fraction of the initial volume (or area) of the ice. Therefore, the choice of lower Prandtl and Schmidt numbers in the simulations is primarily motivated by the need to mitigate the substantial computational expense associated with resolving the thinner boundary layers and longer melt times characteristic of higher Prandtl and Schmidt numbers. In Appendix \ref{appB}, we examine the influence of varying Prandtl and Schmidt numbers in several cases with a low thermal Rayleigh number. While the mean melt rates are significantly lower at higher Prandtl and Schmidt numbers compared to lower values, the non-monotonic dependence of the mean melt rate on ambient salinity and the $\text{Ra}_{S}$ scaling in the high-salinity regime remain unchanged across different Prandtl and Schmidt numbers. Thus, the use of lower Prandtl and Schmidt numbers in simulations does not alter the qualitative findings of the present study.

The initial temperature of the surrounding water is fixed at $\unit{20}{\celsius}$ for all simulations, and the corresponding Stefan number is set to be $\text{Ste}=0.25$. Our simulations cover a parameter range of $\unit{1.5}{\centi\meter} \leq H\leq \unit{30}{\centi\meter}$ and $\unit{0}{\gram\per\kilo\gram}\leq \Delta S\leq \unit{80}{\gram\per\kilo\gram}$, roughly corresponding to $2.2\times 10^{5}\leq \text{Ra}_{T}\leq 1.7\times 10^{9}$ and $0\leq \text{Ra}_{S}\leq 2.3\times 10^{12}$. The values of the height of cylinder $H$ and the corresponding $\text{Ra}_T$, as well as the range of $\Delta S$ and the corresponding $\text{Ra}_S$ are listed in table \ref{tab: tab2}.

\begin{table}
	\begin{center}
		\def~{\hphantom{0}}
\setlength{\tabcolsep}{3mm}
		\begin{tabular}{ccccccc}
			&\unit{H}{(\centi\meter)} & \unit{D}{(\centi\meter)} &$\Delta S$ ($\mathrm{g/kg}$) & $\text{Ra}_{S}$ & $\text{Ra}_{T}$  &  $R_\rho$ \\ [4pt]
                \hline
            2D &    $1.5$ & $0.25$ & $\left [ 0,\,80 \right ]$ & $\left [ 0,\,1.1\times 10^{9} \right ]$  & $2.2\times 10^{5}$ & $[0,52.7]$  \\[4pt]
              2D &  $3$ & $0.5$ & $\left [ 0,\,80 \right ]$ & $\left [ 0,\,9.1\times 10^{9} \right ]$ & $1.7\times 10^{6}$ & $[0,52.7]$\\[4pt]
              2D &  $6$ & $1$ &  $\left [ 0,\,80 \right ]$ & $\left [ 0,\,7.3\times 10^{10} \right ]$ & $1.4\times 10^{7}$ & $[0,52.7]$\\[4pt]
              2D &  $12$ & $2$ &  $\left [ 0,\,80 \right ]$ & $\left [ 0,\,5.8\times 10^{11} \right ]$ & $1.1\times 10^{8}$ & $[0,52.7]$\\[4pt]
               2D & $30$ & $5$ & $\left [ 0,\,20 \right ]$ & $\left [ 0,\,2.3\times 10^{12} \right ]$ & $1.7\times 10^{9}$ & $[0,13.0]$\\[4pt]
               3D &    $1.5$  & $0.25$ &  $\left [ 0,\,80 \right ]$ & $\left [ 0,\,1.1\times 10^{9} \right ]$ &  $2.2\times 10^{5}$ & $[0,52.7]$\\
		\end{tabular}
		\caption{Control parameters used in the simulations, including the values of the height and  diameter of cylinder $H$ and $D$, and the corresponding $\text{Ra}_T$, the range of $\Delta S$ and the corresponding $\text{Ra}_S$ as well as the density ratio $R_\rho$. Note that $\text{Ra}_S=0$ represents the melting in freshwater ($\Delta S = \unit{0}{\gram\per\kilo\gram}$). The notation $\left [ a,\,b \right ]=\left\{x \in \mathbb{R}| a\leq x \leq b \right\}$ denotes a closed interval for $\Delta S$ and $\text{Ra}_S$.}
		\label{tab: tab2}
	\end{center}
\end{table}

Given that the diffusivity of salinity is significantly lower than that of temperature, the salinity field is resolved with much finer spatial resolution compared to that used for the temperature field. Additionally, the phase field is resolved with high precision to accurately capture the thin ice-water interface. Therefore, the multiple-resolution strategy of \citet{Ostilla2015} is applied for both the salinity field and the phase field. For the 2D cases with $H=\unit{1.5}{\centi\meter}$ and $\text{Ra}_T=2.2\times 10^{5}$, a uniform mesh of $n_{x}\times n_{y}=160\times 320$ is employed for the velocity and temperature fields. The uniform mesh for the salinity and phase fields is refined, increasing from $n_{x,f}\times n_{y,f}=480\times 960$ to $960\times 1920$ as $\Delta S$ rises from $\unit{0}{\gram\per\kilo\gram}$ to $\unit{80}{\gram\per\kilo\gram}$. Here, $n_x$ and $n_y$ denote the horizontal and vertical grid resolutions for the velocity and temperature fields, respectively, while $n_{x,f}$ and $n_{y,f}$ represent the refined horizontal and vertical grid resolutions for the salinity and phase fields, respectively. When the height of the ice cylinder $H$ is increased to $\unit{30}{\centi\meter}$, a uniform mesh of $n_x \times n_y = 1920\times 3840$ is employed for the velocity and temperature fields. The refined uniform mesh for the salinity and phase fields ranges from $n_{x,f}\times n_{y,f}=5760\times 11520$ to $9600\times 19200$ as $\Delta S$ increases from $\unit{0}{\gram\per\kilo\gram}$ to $\unit{20}{\gram\per\kilo\gram}$. Due to the high computational cost of the 3D simulations, only seven 3D cases were performed with $\text{Ra}_{T}=2.2\times 10^{5}$ and $\Delta S=\unit{0, 1, 6, 20, 40, 60, 80}{\gram\per\kilo\gram}$ to confirm that the qualitative behaviors of the 3D simulations are similar to those of the 2D simulations. A uniform mesh of $n_x \times n_y \times n_z = 160 \times 320 \times 160$ is used for the velocity and temperature fields, and a finer uniform mesh ranging from $n_{x,f} \times n_{y,f} \times n_{z,f} = 480 \times 960 \times 480$ to $960 \times 1920 \times 960$ is employed for the salinity and phase fields, where $n_z$ and $n_{z,f}$ denote the grid resolution and refined grid resolution in the depth direction, respectively. The grid convergence test of the 2D simulations is shown in Appendix \ref{appA}.

\section{Melt rate}\label{sec:melt_rate}
\begin{figure}\centering
    \includegraphics[width=0.99\linewidth]{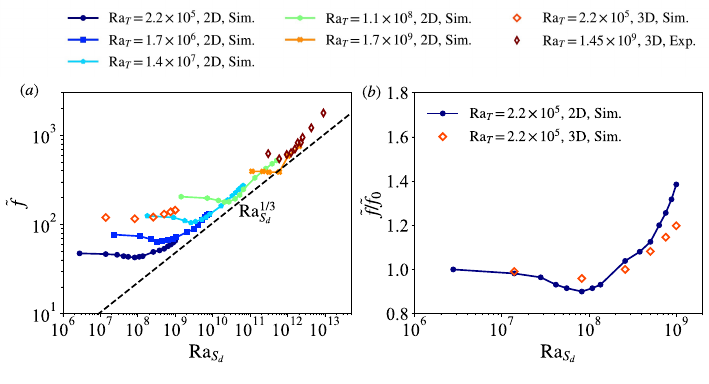}
\caption{(\textit{a}) The normalised mean melt rate $\widetilde{f}$ as a function of $\text{Ra}_{S_d}$ for different $\text{Ra}_{T}$ in simulations as well as the normalised mean melt rate $\widetilde{f}$ as a function of $\text{Ra}_{S}$ at $\text{Ra}_{T}=5.3\times 10^{9}$ in experiments. (\textit{b}) The relative mean melt rate $\widetilde{f}/\widetilde{f}_{0}$ as a function of $\text{Ra}_{S_d}$ at $\text{Ra}_{T}=2.2\times 10^{5}$ in 2D and 3D simulations, where $\widetilde{f}_{0}$ denotes the mean melt rate in freshwater ($\Delta S=\unit{0}{\gram\per\kilo\gram}$) for the respective 2D and 3D cases. Note that $S_i$ is here assumed to be $\unit{0}{\gram\per\kilo\gram}$ for the experiments.}
\label{fig:fig3}
\end{figure}

In this section, we examine the effect of the thermal and saline Rayleigh numbers on the mean melt rate. The normalised mean melt rates $\widetilde{f}$ as a function of $\text{Ra}_{S_d}$ for different $\text{Ra}_T$ in both experiments and simulations are shown in figure \ref{fig:fig3}(\textit{a}). The mean melt rate $\widetilde{f}$ exhibits a non-monotonic relation with $\text{Ra}_{S_d}$ in both experiments and simulations: at constant $\text{Ra}_{T}$, as $\text{Ra}_{S_d}$ increase, the mean melt rate $\widetilde{f}$ initially decreases and then increases again, which is similar to the observations by \citet{Yang2023b}. This non-monotonic behaviour can be attributed to the competition between the thermal and saline buoyancy. At low ambient salinity, the saline buoyancy force is weak, resulting in a predominantly temperature-driven flow. As salinity increases, the saline buoyancy force strengthens, and reaches equilibrium with thermal buoyancy at intermediate $\Delta S$, leading to a reduced melt rate. Further increases in ambient salinity enhance the saline buoyancy force, leading to its dominance at high $\Delta S$. Consequently, the meltwater moves upward more rapidly under the influence of strong saline buoyancy, accelerating the melt rate. 

At high $\text{Ra}_{S_d}$, while the saline boundary layer along the vertical cylinder remains laminar, the bulk flow becomes turbulent, where the mean salt flux follows  $\overline{\text{Nu}}_{S}\propto \text{Ra}_{S_d}^{1/3}$ for the turbulent bulk flow with a laminar boundary layer \citep{bejan1993,Grossmann2000,Grossmann2001,HOLMAN2010}. According to equation (\ref{eqn: NuS}), we have $\widetilde{f} \propto \overline{\text{Nu}}_{S}$. Therefore, the mean melt rate $\widetilde{f}$ should follow $\widetilde{f}\propto \text{Ra}_{S_d}^{1/3}$ at high $\text{Ra}_{S_d}$, which is in agreement with the results shown in figure \ref{fig:fig3}(\textit{a}). The deviation from this trend of the experimental data at high $\text{Ra}_{S_d}$ is most likely due to the variation in temperature between experiments (table \ref{tab: tab1}).

Additionally, some 3D simulations of the vertical ice cylinder are conducted, as depicted in figure \ref{fig:fig3}(\textit{a}). It is found that the mean melt rates $\widetilde{f}$ in the 3D simulations are larger than those in the 2D simulations, primarily due to the larger value of the ratio between the surface area and the volume of the ice in the 3D configuration. The mean melt rate $\widetilde{f}$ in the 3D simulations exhibits a non-monotonic dependence on $\text{Ra}_{S_d}$, consistent with the trends observed in the 2D simulations. At high $\text{Ra}_{S_d}$, the mean melt rate $\widetilde{f}$ appears to transition from a 1/4 scaling to a 1/3 scaling, rather than following a 1/3 scaling as observed in the 2D simulations. This suggests that the 1/3 scaling may emerge at a higher $\text{Ra}_{S_d}$ in the 3D configuration. Nevertheless, the similar non-monotonic behavior observed in both 2D and 3D simulations indicates that the key physical mechanisms governing the melting process are well captured in the 2D simulations.

\begin{figure}\centering
    \includegraphics[width=0.98\linewidth]{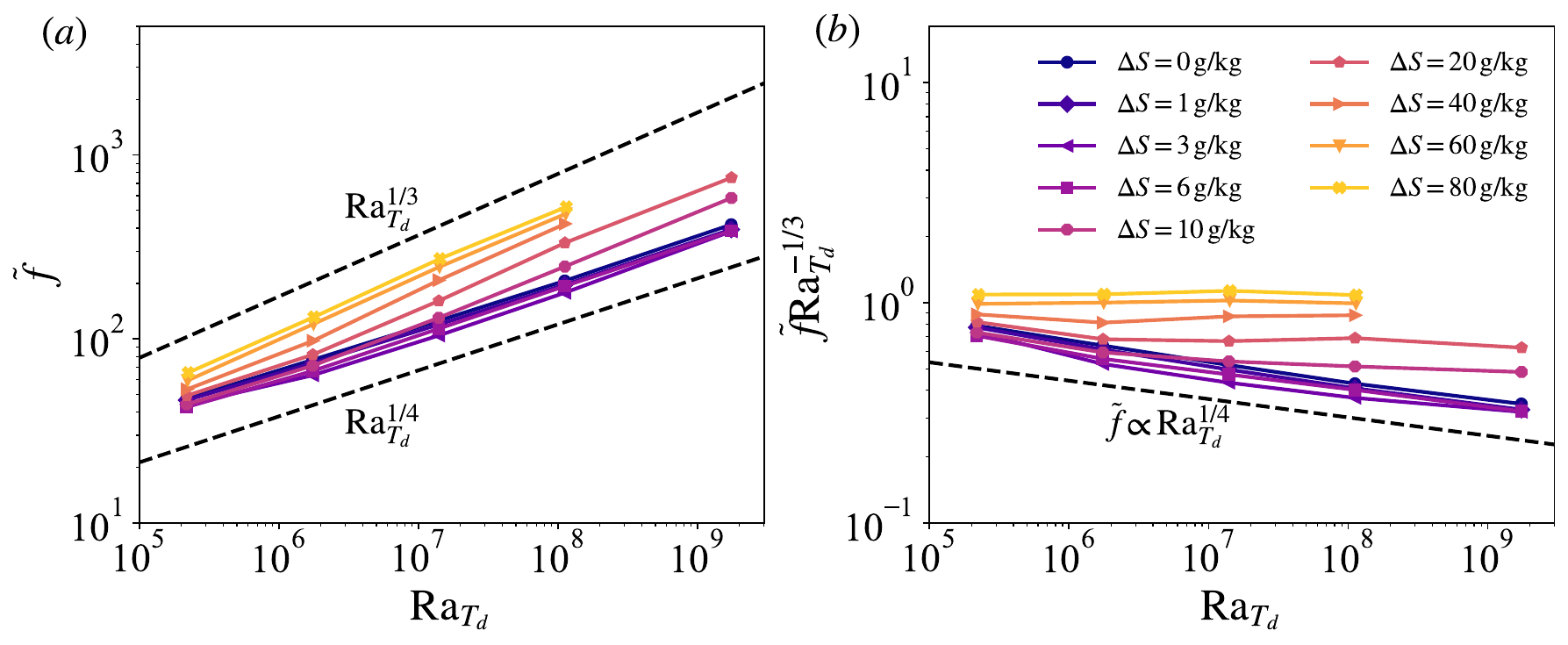}
\caption{(\textit{a}) The normalised mean melt rate $\widetilde{f}$ as a function of $\text{Ra}_{T_d}$} for different $\Delta S$ in simulations. (\textit{b}) The compensated normalised mean melt rate $\widetilde{f}\text{Ra}_{T_d}^{-1/3}$ as a function of $\text{Ra}_{T_d}$ for different $\Delta S$.
\label{fig:fig4}
\end{figure}

\begin{figure}\centering
    \includegraphics[width=0.99\linewidth]{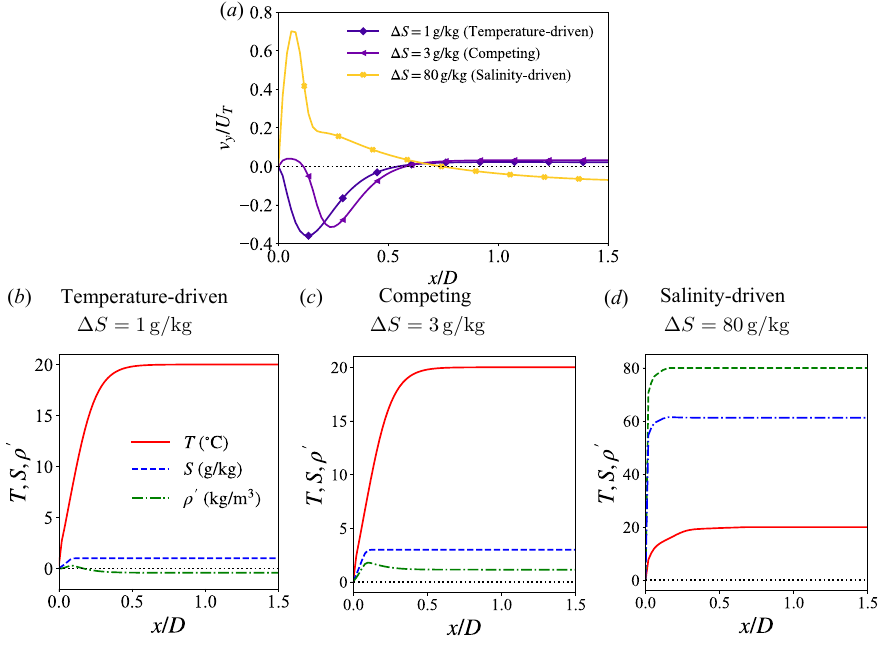}
\caption{(\textit{a}) The normalised time-averaged vertical velocity $v_y/U_T$ as a function of the distance $x/D$ from the melt front at the mid-height of the vertical cylinder in simulations. The velocity is averaged over the time interval $0.2\leq t/t_{m}\leq 0.3$.} Here $U_{T}=\sqrt{g\beta_{T}H\Delta T}$ is the free-fall velocity unit. Three cases with different ambient salinity ($\Delta S=\unit{1}{\gram\per\kilo\gram}$, $\unit{3}{\gram\per\kilo\gram}$, $\unit{80}{\gram\per\kilo\gram}$) and $\text{Ra}_T = 1.4\times 10^{7}$ are shown, with the same colors as in figure \ref{fig:fig4}. (\textit{b}--\textit{d}): The temperature $T$, salinity $S$ and fluctuating density $\rho ^{\prime}$ profiles as a function of the distance $x/D$ from the melt front at different $\Delta S$.
\label{fig:fig5}
\end{figure}

The normalised mean melt rates $\widetilde{f}$ as a function of $\text{Ra}_{T_d}$ for different $\Delta S$ as obtained from our simulations are shown in figure \ref{fig:fig4}. We observe that the normalised mean melt rate $\widetilde{f}$ monotonically increases with increasing $\text{Ra}_{T_d}$ for all $\Delta S$ considered. To gain further insight into the trend of the mean melt rate, the instantaneous vertical velocity $v_{y}$, the temperature $T$, salinity $S$ and fluctuating density $\rho^{\prime}$ profiles in the horizontal direction at the mid-height of the vertical cylinder are shown in figure \ref{fig:fig5}. The buoyancy forces induced by temperature and salinity act downward and upward, respectively, corresponding to a higher density at the wall compared to the ambient for temperature-driven flow (figure \ref{fig:fig5}\textit{b}) and a lower density at the wall compared to the ambient for salinity-driven flow (figure \ref{fig:fig5}\textit{d}).

When the ambient salinity is small, the temperature dependence dominates the buoyancy, causing the cold meltwater to descend, as indicated by the instantaneous vertical velocity $v_{y}$ in figure \ref{fig:fig5}(\textit{a}). This results in the formation of a laminar thermal boundary layer along the vertical ice cylinder within the laminar bulk flow, with the mean heat flux following the $\overline{\text{Nu}}_{T}\propto \text{Ra}_{T_d}^{1/4}$ scaling for the laminar bulk flow with a laminar boundary layer \citep{bejan1993,Grossmann2000,Grossmann2001,HOLMAN2010}. According to equation (\ref{eqn: NuT}), we have $\widetilde{f} \propto \overline{\text{Nu}}_{T}$. Consequently, the mean melt rate $\widetilde{f}$ adheres to the $\widetilde{f} \propto \text{Ra}_{T_d}^{1/4}$ scaling relation for low $\Delta S$ values ($\Delta S < \unit{2}{\gram\per\kilo\gram}$), which is consistent with the observations seen in figure \ref{fig:fig4}.

At intermediate $\Delta S$, the buoyancy forces driven by temperature and salinity counteract, resulting in a bi-directional flow near the melt front, as depicted in figure \ref{fig:fig5}(\textit{a}). Due to the faster diffusion of heat compared to salt ($\text{Le}=100$), the saline boundary layer is much thinner than the thermal boundary layer. Within the saline boundary layer, cold and fresh meltwater ascends due to salinity dominated buoyancy, while cold and saline ambient water descends due to thermal buoyancy outside the saline boundary layer and within the thermal boundary layer. This competition between thermal and saline buoyancies attenuates the buoyancy-driven flow and vertical velocity, consequently decelerating the melt rate. The flow around the vertical ice cylinder remains laminar, with the mean heat flux adhering to $\overline{\text{Nu}}_{T}\propto \text{Ra}_{T_d}^{1/4}$\citep{bejan1993,Grossmann2000,Grossmann2001,HOLMAN2010}. Accordingly, the mean melt rate $\widetilde{f}$ should also follow a $\widetilde{f}\propto \text{Ra}_{T_d}^{1/4}$ scaling for intermediate $\Delta S$ ($\unit{2}{\gram\per\kilo\gram} \leq \Delta S \leq \unit{10}{\gram\per\kilo\gram}$), consistent with the numerical findings presented in figure \ref{fig:fig4}.

However, when the ambient salinity is large, salinity dominates the buoyancy, causing the fresh meltwater to ascend, as depicted by the instantaneous vertical velocity $v_{y}$ in figure \ref{fig:fig5}(\textit{a}). Notably, figure \ref{fig:fig5}(\textit{a}) reveals that near the melt front ($x/D\approx 0.06$), the magnitude of the positive $v_{y}$ in the case with $\Delta S=\unit{80}{\gram\per\kilo\gram}$ is nearly twice as large as those observed in the cases with $\Delta S=\unit{1}{\gram\per\kilo\gram}$ and $\unit{3}{\gram\per\kilo\gram}$. This increase can be attributed to the larger difference in density between the meltwater and the ambient water (figure \ref{fig:fig5}\textit{d}), resulting in accelerated upward motion of the meltwater with increasing ambient salinity. The intrusion of warm ambient water onto the melt front occurs more rapidly, amplifying the melt rate and surface heat flux. With increasing $\Delta S$, the surrounding boundary layer remains laminar, while the bulk flow begins to transition from laminar to turbulent, resulting in a transition in the scaling of the mean melt rate. Figure \ref{fig:fig4} shows that this shift in scaling of the mean melt rate towards $\widetilde{f} \propto \text{Ra}_{T_d}^{1/3}$ for high ambient salinity ($\Delta S > \unit{10}{\gram\per\kilo\gram}$), which is consistent with the mean heat flux scaling $\overline{\text{Nu}}_{T}\propto \text{Ra}_{T_d}^{1/3}$ for a turbulent bulk flow with a laminar boundary layer \citep{bejan1993,Grossmann2000,Grossmann2001,HOLMAN2010}.

The $\text{Ra}_S$-scaling at high ambient salinity is analogous to the $\mathscr{V} \propto \text{Gr}^{1/3}=(\text{Ra}/\text{Pr})^{1/3}$ scaling derived by \citet{Mondal2019}, where $\mathscr{V}$ is the ablation velocity of a vertical ice face in cold seawater. In addition, the $\mathscr{V} \propto \Delta T_d^{4/3}$ relation for a vertical ice wall in cold seawater reported by \citet{Kerr2015} and \citet{Gayen2016} essentially represents a heat flux scaling of $\Delta T_d^{1/3}$, similar to the $\text{Ra}_{T_d}^{1/3}$ scaling that we observe here for high $\text{Ra}_S$. The similarity between the scalings reported by these studies for cold ambient water and those presented here for warm ambient water suggests that the physics in both cases is comparable, provided that the relative contributions of temperature and salinity on the buoyancy are the same. 

\section{Flow regimes}\label{sec:flow_regimes}

\begin{figure}\centering
    \includegraphics[width=0.99\linewidth]{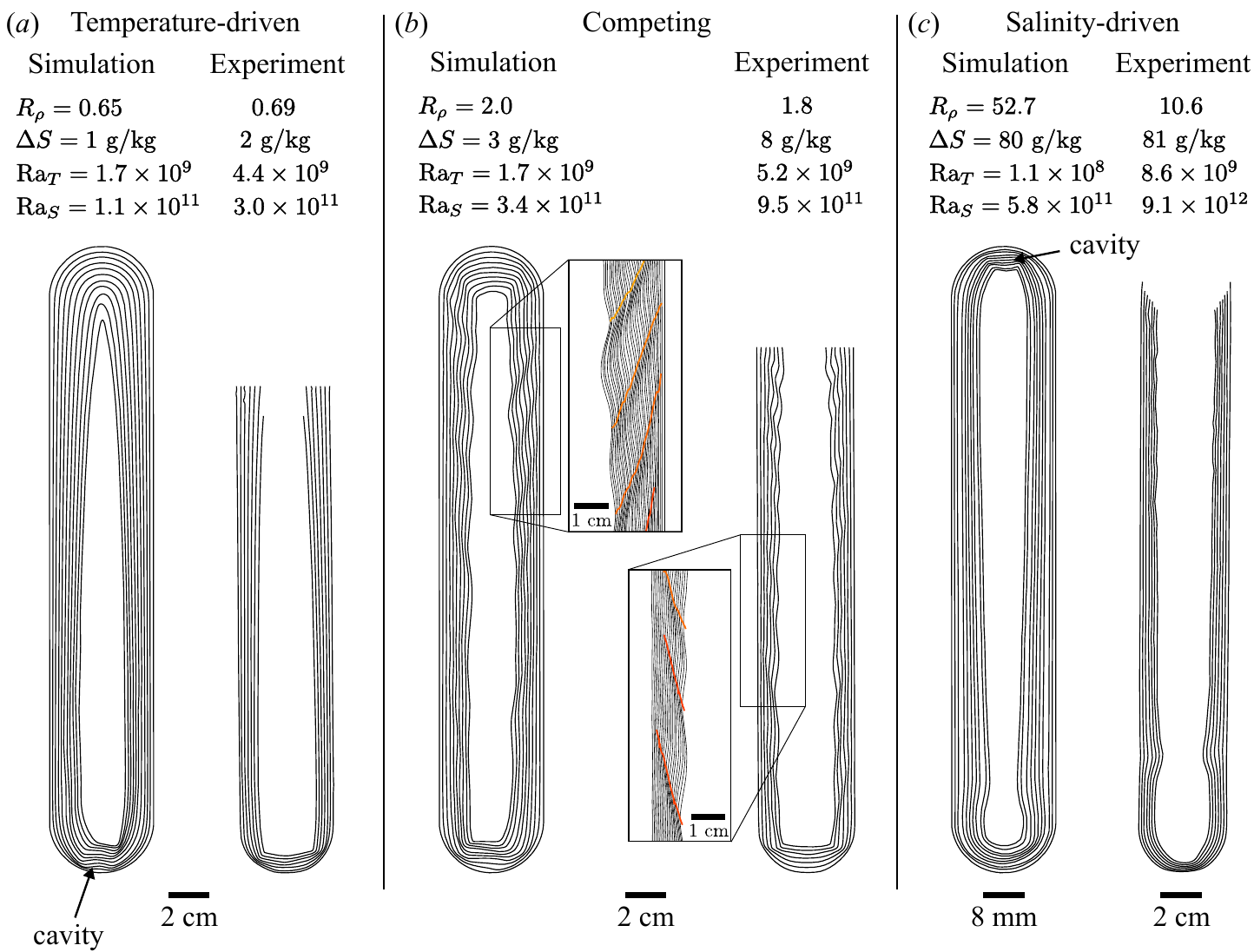}
\caption{Morphological evolution of the ice cylinder in simulations (left in each panel) and experiments (right in each panel) for the three different flow regimes: (\textit{a}) temperature-driven flow,  (\textit{b}) competing flow, and (\textit{c}) salinity-driven flow. The time intervals between the contours from simulations and experiments are $0.1 t_m$ and $0.15 t_m$, respectively, where $t_m$ is the time needed to melt 70\% of the initial area in 2D simulations and the volume in 3D experiments. The innermost contours correspond to $t = t_m$.  The insets in (\textit{b}) show an enlarged part of the contours, with the tracked position of the crests represented by red lines.} In the experiments not the entire cylinder is imaged and thus only the bottom part of each cylinder is visible.
\label{fig: fig6}
\end{figure}

In this section, we investigate the influence of ambient salinity on the evolution of ice surface morphology. As discussed previously, the flow patterns can be divided into three regimes: temperature-driven, competing, and salinity-driven flow. These flow regimes can be quantified using the density ratio $R_\rho$, defined in equation (\ref{eqn:Le_Pr_Rp}). The limits $R_\rho \ll 1$ and $R_\rho \gg 1$ correspond to fully temperature-driven and fully salinity-driven flow, respectively. Using both experiments and simulations, we here explore the ice surface morphology dynamics for different $R_\rho$, by varying the ambient salinity while keeping the ambient temperature constant.

\begin{figure}\centering
    \includegraphics[width=0.98\linewidth]{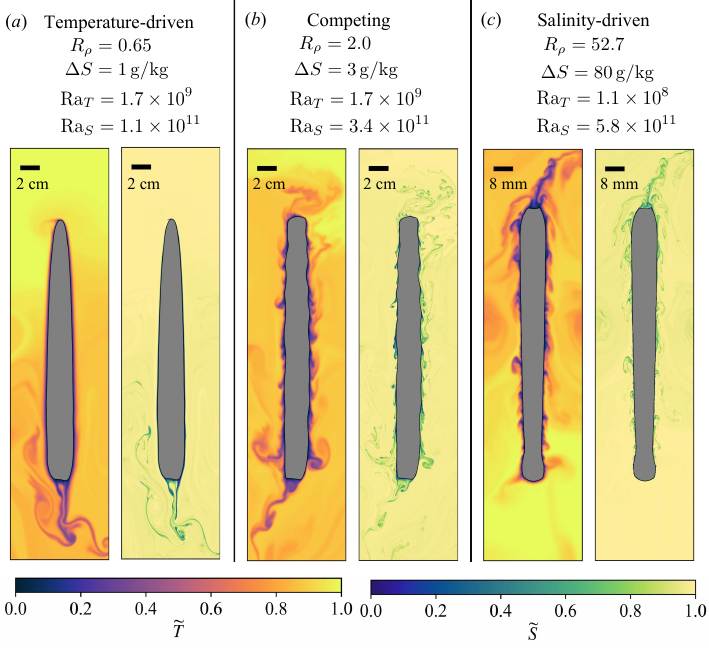}
\caption{Snapshots of the temperature (left) and salinity (right) fields for the three different flow regimes in the simulations at time $t/t_{m}=0.8$: (\textit{a}) temperature-driven flow,  (\textit{b}) competing flow, and (\textit{c}) salinity-driven flow. Here $\widetilde{T}=T/\Delta T$ and $\widetilde{S}=S/\Delta S$ .}
\label{fig: fig7}
\end{figure}

Both our experiments and simulations show that the evolution of the ice surface morphology depends heavily on $R_\rho$, exhibiting distinct dynamics for each of the three flow regimes (figure \ref{fig: fig6}). Remarkably, the dynamics of the morphology from the 2D simulations closely resembles those obtained from 3D experiments at similar $R_\rho$. In the following paragraphs, we discuss the evolution of the ice morphology for each flow regime separately, and relate it to the corresponding flow, based on the instantaneous temperature and salinity fields obtained from simulations (figure \ref{fig: fig7}).


For temperature-driven flow ($R_\rho < 1$), we found that the ice cylinder exhibits a pointed top and a flat bottom (figure \ref{fig: fig6}\textit{a}). The denser cold meltwater adjacent to the cylinder descends along its surface, insulating the lower part of the cylinder while exposing the upper part to the warmer ambient water. This results in a higher melt rate at the top as compared to the bottom, leading to the observed morphological dynamics. The difference in melt rate between the top and the bottom is obvious in the contours from the simulation (left in figure \ref{fig: fig6}\textit{a}), but less so for the experiment (right in figure \ref{fig: fig6}\textit{a}). This is possibly due to the difference in shape of the top of the cylinder, which is round in the simulation, but flat in the experiment (for practical reasons). As this will only affect a downward flow, it will mainly be relevant in the temperature-driven regime. Additionally, a cavity is formed at the bottom of the ice cylinder in the simulation (left in figure \ref{fig: fig6}\textit{a} for times $0.1 \leq t/t_m \leq 0.8$). This cavity formation can be attributed to flow separation at the bottom of the cylinder, which generates attached convective rolls. These convection rolls, evident from the instantaneous salinity field (figure \ref{fig: fig7}\textit{a}), enhance local mixing, increasing the local heat flux and therewith the melt rate, ultimately leading to the formation of the cavity. Around $t = 0.8 t_m$, the diameter of the cylinder is almost half its initial value, such that the convection rolls are disturbed by the downward flow along the cylinder, causing them, and therefore the cavity, to disappear. In the experimental contours (right in figure \ref{fig: fig6}\textit{a}), the cavity is not visible due to the optical limitations of the side view. 

For competing flow ($R_\rho \approx 2$), a regular scallop pattern appears on the side of the cylinder (figure \ref{fig: fig6}\textit{b}). Between the downward-flowing layer of cold ambient water and the ice surface, a thin layer of fresh meltwater flows upward along the side of the cylinder. This bi-directional flow is clearly evident from the instantaneous vertical velocity profile for $\Delta S=\unit{3}{\gram\per\kilo\gram}$ (figure \ref{fig:fig5}\textit{a}). The shear between the upward- and downward-flowing layers induces a Kelvin--Helmholtz-type instability, akin to that observed by \citet{Weady2022} when they melted an ice cylinder in $\unit{5.6}{\celsius}$ freshwater. Similar to the case of \citet{Weady2022}, our results show that the scallop pattern is carved by an unstable bidirectional flow. In their case the flow consists of a positively buoyant inner layer of water below the maximum density temperature, nested inside a negatively buoyant outer layer of water above the maximum density temperature. In our case, the positively buoyant inner layer is cold and fresh meltwater, and the negatively buoyant outer layer is cold and saline ambient water. Due the difference in salinity between the layers, the density difference is 10 to 100 times larger than in the freshwater case of \citet{Weady2022}. In addition, the widths of the inner and outer layers are controlled by the saline and thermal diffusivity, respectively, while only the latter is relevant in freshwater. Despite these differences, the resulting scallop pattern observed in our saline water cases is qualitatively similar to the freshwater case of \citet{Weady2022}. We conclude that scallop patterns emerge due to the presence of a bidirectional flow, regardless of the nature of this flow. The characteristics of these scallops are discussed in more detail in § \ref{sec:morphology}.

In salinity-driven flow ($R_\rho \gg 1$), the ice melts fastest just above the bottom of the cylinder, narrowing the cylinder's width at this point (figure \ref{fig: fig6}\textit{c}). Above this point, cold meltwater ascends along the cylinder, shielding the ice from warm ambient water and consequently reducing its melting rate. Below this point, thermal driving is strong enough to prevent meltwater from flowing up along the cylinder. The cold and fresh meltwater ascends within the saline boundary layer, while the cold and saline ambient water descends outside the saline boundary layer but within the thermal boundary layer, as confirmed by the instantaneous snapshots of temperature and salinity fields (figure \ref{fig: fig7}\textit{c}). This bi-directional flow shields the melt front from direct contact with the warm ambient water, leading to a slower local melt rate at the bottom region. At the point of minimum width, the melt rate increases due to the intrusion of warm ambient water, as cold meltwater either flows up along the cylinder or is advected downward by the temperature-driven flow. This observation is similar to the bifurcation point observed by \citet{Josberger1981} in their experiments on a vertical ice wall. Furthermore, similar to the cavity formed in the cylinder with temperature-driven flow (figure \ref{fig: fig6}\textit{a}), a cavity is formed with salinity-driven flow (figure \ref{fig: fig6}\textit{c}), but at the top of the cylinder, instead of at the bottom. Here, the upward flow of meltwater separates from the ice surface at the top, generating convection rolls that enhance heat transport to the ice and thus the melt rate.

\begin{figure}\centering
    \includegraphics[width=\linewidth]{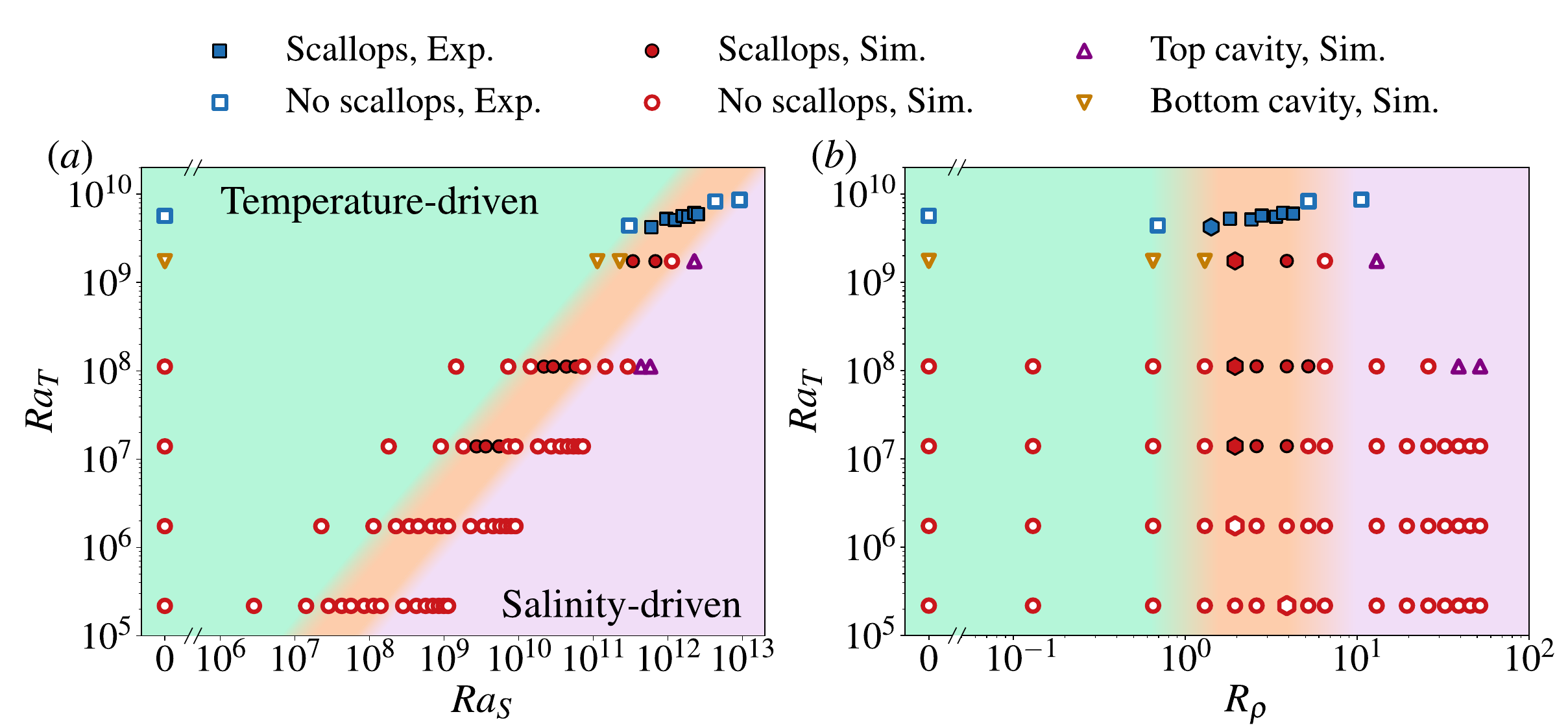}
\caption{The (\textit{a}) $\text{Ra}_{T}$-$\text{Ra}_{S}$ and (\textit{b}) $\text{Ra}_{T}$-$R_{\rho}$ phase diagrams for various morphological patterns observed in simulations and experiments. The green, orange, and purple regions correspond to the temperature-driven, competing, and salinity-driven flow regimes, respectively. The large hexagonal markers in (\textit{b}) indicate the minimum mean melt rate case for each $\text{Ra}_T$, as discussed in § \ref{sec:melt_rate}.}
\label{fig: fig8}
\end{figure}

The extent of these three flow regimes and their implications on the ice morphology are summarized in the $\text{Ra}_{T}$-$\text{Ra}_{S}$ and $\text{Ra}_{T}$-$R_{\rho}$ phase diagrams in figures \ref{fig: fig8}(\textit{a}) and \ref{fig: fig8}(\textit{b}), respectively. The boundary between the temperature-driven and competing regimes is defined at $R_\rho = 1$, corresponding to the zero wall-shear condition established by \citet{Carey1982}. For the boundary between the competing and salinity-driven regimes, we adopt $R_\rho = 6.37$, corresponding to the density ratio for which \citet{Josberger1981} found the wall-normal integral of buoyancy near a vertical ice wall to be zero. It should be noted that these values of $R_\rho$ serve as indicative rather than strict boundaries, as the transitions between the regimes are continuous. Nevertheless, our observations of scallops in both experiments and simulations align closely with the competing flow regime (figure \ref{fig: fig8}). Scallop patterns are only observed for $\text{Ra}_T \gtrsim 10^7$, indicating a minimum thermal driving for triggering an instability. Bottom and top cavities only appear at sufficiently high $\text{Ra}_T$ and $\text{Ra}_S$, respectively, which is likely linked to a critical Reynolds number and thus a minimal Rayleigh number for which flow separation occurs.


\section{Properties of the ice scallops}\label{sec:morphology}
\begin{figure}\centering
    \includegraphics[width=0.7\linewidth]{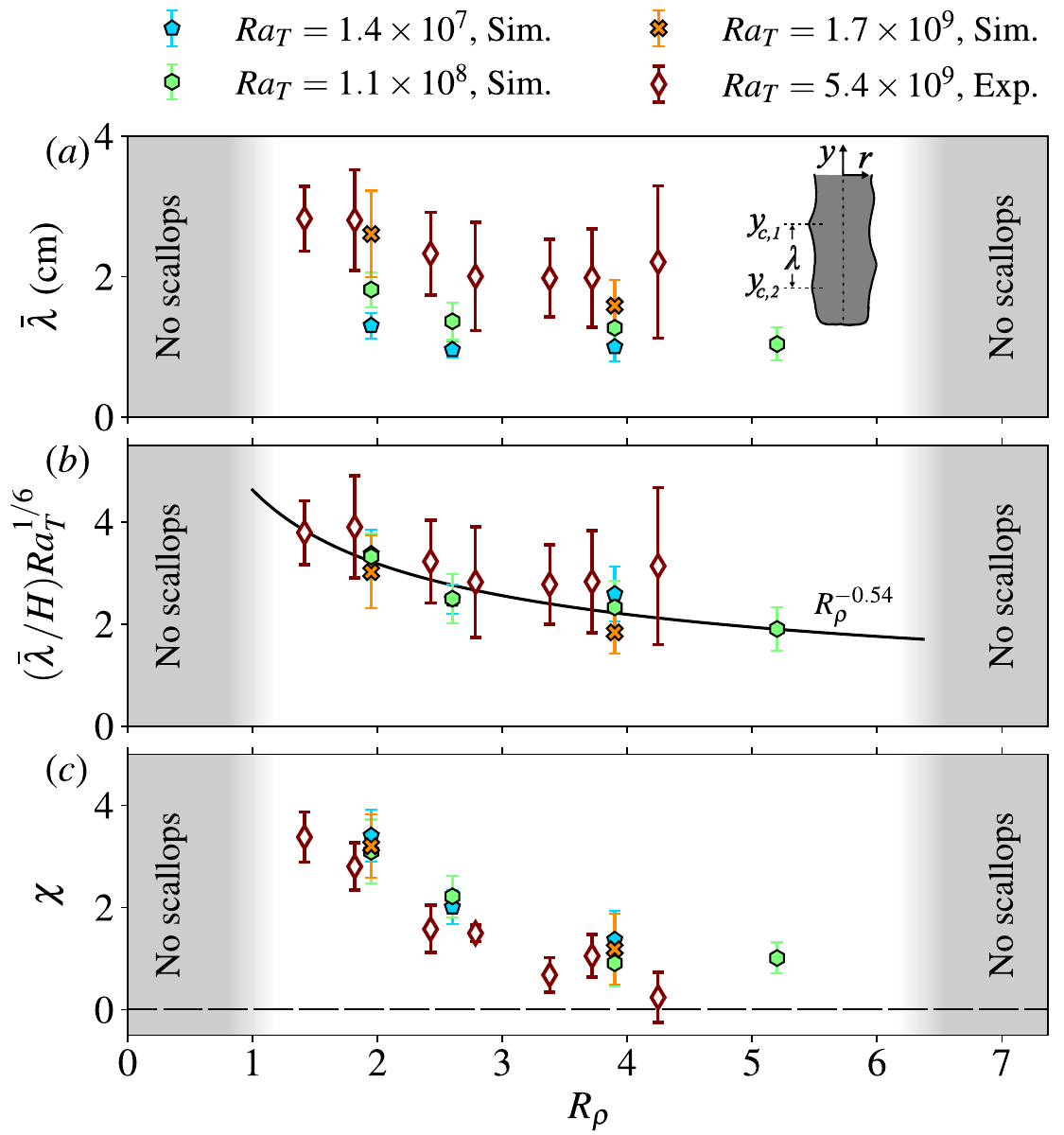}
\caption{(\textit{a}) The mean wavelengths $\bar{\lambda }$, (\textit{b}) the compensated mean wavelengths $ \left ( \bar{\lambda } /H \right )\text{Ra}_{T}^{1/6}$ and (\textit{c}) the scallop migration velocities $\chi$ as a function of the density ratio $R_{\rho }$ in simulations and experiments. Each point corresponds to a single simulation or experiment. The interval markers represent one standard deviation spread in wavelength or migration velocity. The inset in (\textit{a}) shows the definition of the wavelength $\lambda$ and the coordinate system for computation of $\chi$ (equation (\ref{eq: chi})).}
\label{fig: fig9}
\end{figure}

In this section, we discuss the scallop patterns observed in the thermal-saline competing regime (figures \ref{fig: fig6}\textit{b} and \ref{fig: fig7}\textit{b}). The typical length scale for the scallop patterns, wavelength $\lambda$, obviously depends on the ambient conditions. To investigate the dependence of the scallop wavelength on the ambient conditions, we compute the mean (in space and time) of all scallop wavelengths $\bar{\lambda}$ for each case. The mean wavelengths $\bar{\lambda }$ as a function of the density ratio $R_{\rho }$ in simulations and experiments are depicted in figure \ref{fig: fig9}(\textit{a}). As shown by the data from simulations, the mean wavelength $\bar{\lambda }$ increases with increasing $\text{Ra}_T$ at similar $R_\rho$, indicating that the thermal driving strength influences the scallop pattern. At similar $\text{Ra}_T$, the wavelength observed at low $R_\rho$ (low salinity) is up to twice the value observed for high $R_\rho$ (high salinity). Yet, no clear correlation between $\bar{\lambda }$ and $R_\rho$ emerges, given that the standard deviation spread in wavelengths exceeds 10\% for each case, as indicated by the interval markers in figure \ref{fig: fig9}(\textit{a}). The mean wavelengths obtained from the simulations with $\text{Ra}_T = 1.7 \times 10^9$ agree well with those observed in experiments performed at similar ambient conditions ($\text{Ra}_T = 5.3 \times 10^9$). 

To further explore the dependence of $\bar{\lambda }$ on $\text{Ra}_T$ and $R_\rho$, we perform a least-squares fit of the form $\bar{\lambda }/H = a \text{Ra}_T^b R_\rho^c$, where $H$ is the cylinder height and $a$, $b$, and $c$ are the free parameters. It yields the empirical relation:

\begin{equation}
    \frac{\bar{\lambda }}{H} = (3.8 \pm 1.8) \; \text{Ra}_T^{-0.16 \pm 0.02}\, \; R_\rho^{-0.54 \pm 0.15},
\end{equation}
where the $\pm$-values correspond to standard errors.



A possible explanation for this scaling of $\bar{\lambda }$ with $\text{Ra}_T$ can be given by constructing a typical length scale for the temperature-driven flow. For a vertical cylinder in a single-phase fluid with $\text{Pr} \geq 1$, the velocity $U$ of a thermally-driven flow follows $U \propto (\kappa_T/H) \text{Ra}_T^{1/2}$ \citep{bejan1993}, with the local buoyancy timescale $\tau = (\nu/(g\beta_T\Delta T)^2)^{1/3}$. Assuming that the mean wavelength is proportional to the typical length scale of the flow $U\tau $, we obtain:
\begin{equation}
    \bar{\lambda } \propto  U \tau = \frac{\kappa_T}{H} \text{Ra}_T^{1/2} \left(\frac{\nu}{(g\beta_T\Delta T)^2}\right)^{1/3} = H\text{Ra}_T^{-1/6}.
    \label{eqn: length_scale}
\end{equation}

The above scaling can be evaluated using the compensated mean wavelengths $ \left ( \bar{\lambda } /H \right )\text{Ra}_{T}^{1/6}$, as shown in figure \ref{fig: fig9}(\textit{b}). We find that this provides a better collapse of the mean wavelengths for the parameter space explored in this work. Moreover, the exponent obtained from the fit ($\text{Ra}_T^{-0.16 \pm 0.02}$) is in good agreement with the scaling $\bar{\lambda}/H \propto \text{Ra}_T^{-1/6}$ in equation (\ref{eqn: length_scale}). Given the large spread in wavelengths, this scaling clearly warrants further investigation. Specifically, the differences and interactions between individual scallops should be explored. Scallop wavelengths could depend on their vertical position \citep{Weady2022}, and neighbouring scallops could merge or split over time. Such details are not accessible when a mean wavelength is used. Although a robust theory for the pattern formation therefore remains elusive, we infer that both the driving strength and the ratio of thermal to saline driving play important roles in determining the scallop wavelength, underscoring their significance in future investigations.


Another notable feature of the scallop pattern is that it migrates downward over time (figure \ref{fig: fig6}\textit{b}). Similar migration of scallops has been observed in horizontal blocks of dissolving plaster and melting ice, and inclined blocks of dissolving caramel, salt, and plaster, subject to forced convection \citep{Blumberg1974, Gilpin1980, Bushuk2019, Cohen2020}. 

To quantify the downward migration, we define a dimensionless migration velocity $\chi$ as the vertical movement of the scallop pattern compared to its horizontal movement. The vertical speed is given by the change in vertical position $y_c(t)$ of scallop crests (maxima in local radius) over time ($d y_c/d t$), and the horizontal speed is given by the height-averaged horizontal melt rate ($\left\langle \partial  r/\partial  t \right \rangle_{y}$), leading to the following definition of the migration velocity $\chi$:

\begin{equation}
   \chi=\frac{\left\langle \frac{d{y}_c}{d t}\right\rangle_{t, N}}{\left \langle \frac{\partial  r(y, t)}{\partial  t} \right \rangle_{y, t}},
   \label{eq: chi}
\end{equation}

\begin{figure}\centering
    \includegraphics[width=0.7\linewidth]{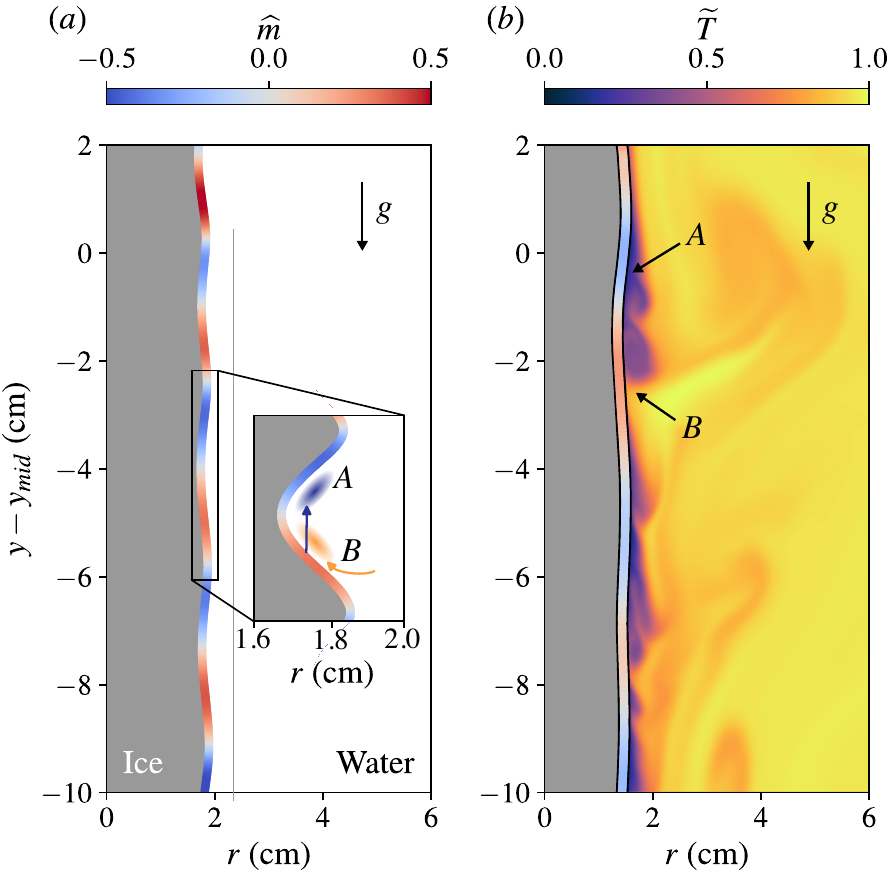}
\caption{(\textit{a}) Normalised horizontal melt rate along the contour of the cylinder from the experiment at $R_\rho = 1.4$ ($\text{Ra}_T = 4.2\times 10^{9}$, $\text{Ra}_S = 6.0 \times 10^{11}$), averaged over \unit{10}{\min}. Positive and negative values indicate above average and below average melt rates, respectively. (\textit{b}) Snapshot of the temperature field from the numerical simulation at $R_\rho = 2.0$ ($\text{Ra}_T = 1.7 \times 10^9$, $\text{Ra}_S = 3.4 \times 10^{11}$), with local melt rates on the contour, averaged over 2 minutes. Points A and B indicate the accumulation of cold meltwater and intrusion of warm ambient water, respectively, also sketched in the inset of (\textit{a}).}
\label{fig: fig10}
\end{figure}

\noindent where the subscripts $y$, $t$, and $N$ indicate averages over height, time, and number of scallop crests, respectively. The coordinate system and crest locations are sketched in the inset of figure \ref{fig: fig9}(\textit{a}). \citet{Blumberg1974} and \citet{Bushuk2019} observed a similar pattern migration in their experiments and quantified this movement using a crest propagation angle $\phi$. The migration velocity $\chi$ (equation (\ref{eq: chi})) is related to this propagation angle as $\chi = \cot(\phi)$. With increasing $R_\rho$, the migration velocity decreases monotonically (figure \ref{fig: fig9}\textit{c}). The average migration velocities from simulations agree well with those observed in experiments, further supporting the suitability of the 2D simulations in representing the 3D experiments. There is little variation between different $\text{Ra}_T$, thus $\chi$ is nearly independent of the thermal driving strength.  

To explain the observed migration, we show the normalised local melt rate along the ice contour for the experiment at $R_\rho = 1.4$ in figure \ref{fig: fig10}(\textit{a}). It is defined as $\hat{m} = \left (\langle \dot{r} \rangle_t - \langle \dot{r} \rangle_{y, t} \right ) / \langle \dot{r} \rangle_{y, t}$, where the subscripts $y$ and $t$ indicate averages over height and time, respectively. For the computation of $\langle \dot{r} \rangle_t$, the horizontal melt rate profile at each time $t$ was shifted vertically by a distance $-\langle dy_c/dt\rangle_{t,N} (t-t_0)$ to account for the scallop migration since a time $t_0$. The melt rates $\hat{m}$ shown in figure \ref{fig: fig10} are therefore relative to the scallop position. As shown in figure \ref{fig: fig10}(\textit{a}), the melt rate is larger at the upper side of each crest compared to its lower side. This asymmetry leads to faster ablation of the ice on the upper side of the crests, resulting in the apparent downward migration of the pattern. 

To explore the origin of this asymmetric melt rate, we show the instantaneous temperature field from the simulation with $R_\rho = 2.0$ in figure \ref{fig: fig10}(\textit{b}). A notable temperature disparity is evident between the two sides of the crests, resulting in an uneven melt rate. Between adjacent crests, cold meltwater moves up along the ice and accumulates on the lower side of a crest (points A in figure \ref{fig: fig10}\textit{a}--\textit{b}). This displacement of fluid allows warm ambient water to intrude (points B in figure \ref{fig: fig10}\textit{a}--\textit{b}). 



Using this proposed mechanism for pattern migration, we can explain our observations from figure \ref{fig: fig9}(\textit{c}). With increasing $R_\rho$, the saline buoyancy of the meltwater increases, such that less meltwater accumulates behind the scallop crests and instead moves upward past the crests. This reduces the local differences in melt rate and therefore decreases the migration velocity, explaining the decrease of $\chi$ with increasing $R_\rho$ in figure \ref{fig: fig9}(\textit{c}).

\section{Conclusions}\label{sec:conclusion}
In this study, we examined the melting process of a vertical ice cylinder in saline water through a combination of laboratory experiments and direct numerical simulations. Depending on the density ratio $R_{\rho}$, controlled by varying the ambient salinity, the flow can be categorized into three flow regimes: temperature-driven flow, salinity-driven flow and thermal-saline competing flow. We propose scaling laws for the mean melt rate within these regimes and report distinct morphology dynamics for each flow regime.

We find that the mean melt rate exhibits a non-monotonic relationship with ambient salinity: as the salinity increases, the mean melt rate initially decreases towards the point that thermal and saline buoyancies compensate each other, and then increases again. The slowest mean melt rate appears in the competing flow regime. Moreover, the mean melt rate monotonically increases with increasing thermal Rayleigh number. In the temperature-driven and competing flow regimes, the mean melt rate follows a $\widetilde{f}\propto \text{Ra}_{T_d}^{1/4}$ scaling, consistent with the mean heat flux scaling law for the laminar bulk flow with a laminar boundary layer \citep{bejan1993,Grossmann2000,Grossmann2001,HOLMAN2010}. However, a scaling transition occurs when the ambient salinity is large. In the salinity-driven flow regime, the mean melt rate transitions to a $\widetilde{f}\propto \text{Ra}_{T_d}^{1/3}$ scaling, adhering to the mean heat flux scaling law for the turbulent bulk flow with a laminar boundary layer \citep{bejan1993,Grossmann2000,Grossmann2001,HOLMAN2010}. Furthermore, the mean melt rate also follows a $\widetilde{f}\propto \text{Ra}_{S_d}^{1/3}$ scaling in this regime, consistent with the mean salt flux scaling law for the turbulent bulk flow with a laminar boundary layer \citep{bejan1993,Grossmann2000,Grossmann2001,HOLMAN2010}.

The ice cylinder exhibits distinct morphology dynamics across different flow regimes. In the temperature-driven flow regime, the ice cylinder sharpens at the top, while remaining flat at the bottom. This morphology arises from the downward movement of cold meltwater driven by thermal buoyancy. When the thermal Rayleigh number is sufficiently high, a cavity forms at the bottom of the ice cylinder due to flow separation, leading to enhanced local mixing that increases the local melt rate. In the salinity-driven flow regime, the minimum width appears near the bottom of the cylinder. This is caused by cold meltwater moving upward, driven by saline buoyancy. When the saline Rayleigh number is sufficiently high, a cavity forms at the top of the ice cylinder. In the thermal-saline competing flow regime, distinctive scallop patterns emerge along the side of the ice cylinder, resulting from the competition between thermal and saline buoyancies. The wave crests of these scallop patterns migrate downward over time, due to a difference in melt rate just above and below the crests. We find that this phenomenon is caused by the simultaneous accumulation of meltwater on the lower side of the crests and depletion of meltwater on the upper side of the crests.

Our results elucidate the effect of ambient salinity on the melting process of an ice cylinder in saline water. The proposed scaling laws for the mean melt rate may improve predictions of subaqueous melting for tidewater glaciers and icebergs in geophysical settings characterized by low to moderate Rayleigh numbers, where both thermal and saline boundary layers remain laminar. However, such systems are often characterized by significantly higher Rayleigh numbers, for which the boundary layers may transition from laminar to turbulent. The scaling laws under such conditions remain to be established. For realistic Schmidt numbers ($\mathcal{O}(1000)$), the transition is expected to occur in a range around $Ra_{S_d} = \mathcal{O}(10^{20})$ \citep{Lohse2024}. The experiments and simulations provide an initial insight into the ambient conditions under which morphological features like scallops and cavities appear on the surface of an ice cylinder. However, the onset for the emergence of scallop patterns remains unclear and warrants further investigation. 

The next step is to extend our experimental and numerical results to the case of colder ambient water ($T_{\infty }<\unit{5}{\celsius}$), which is more relevant in the geophysical context. Then obviously the dependence of the density on the temperature and salinity becomes more complex, but it is well known, so that a comparison between experiments and simulations should again be possible.

\backsection[Funding]{We thank Gert-Wim Bruggert, Martin Bos, and Thomas Zijlstra for technical support. This work was financially supported by the European Union (ERC, MeltDyn, 101040254 and ERC, MultiMelt, 101094492). We also acknowledge the EuroHPC Joint Undertaking for awarding the projects EHPC-REG-2022R03-208 and EHPC-REG-2023R03-178 access to the EuroHPC supercomputer Discoverer, hosted by Sofia Tech Park (Bulgaria).}

\backsection[Declaration of interests]{The authors report no conflict of interest.}

\FloatBarrier
\appendix
\section{The grid convergence test of the simulations}\label{appA}
\begin{figure}\centering
    \includegraphics[width=0.98\linewidth]{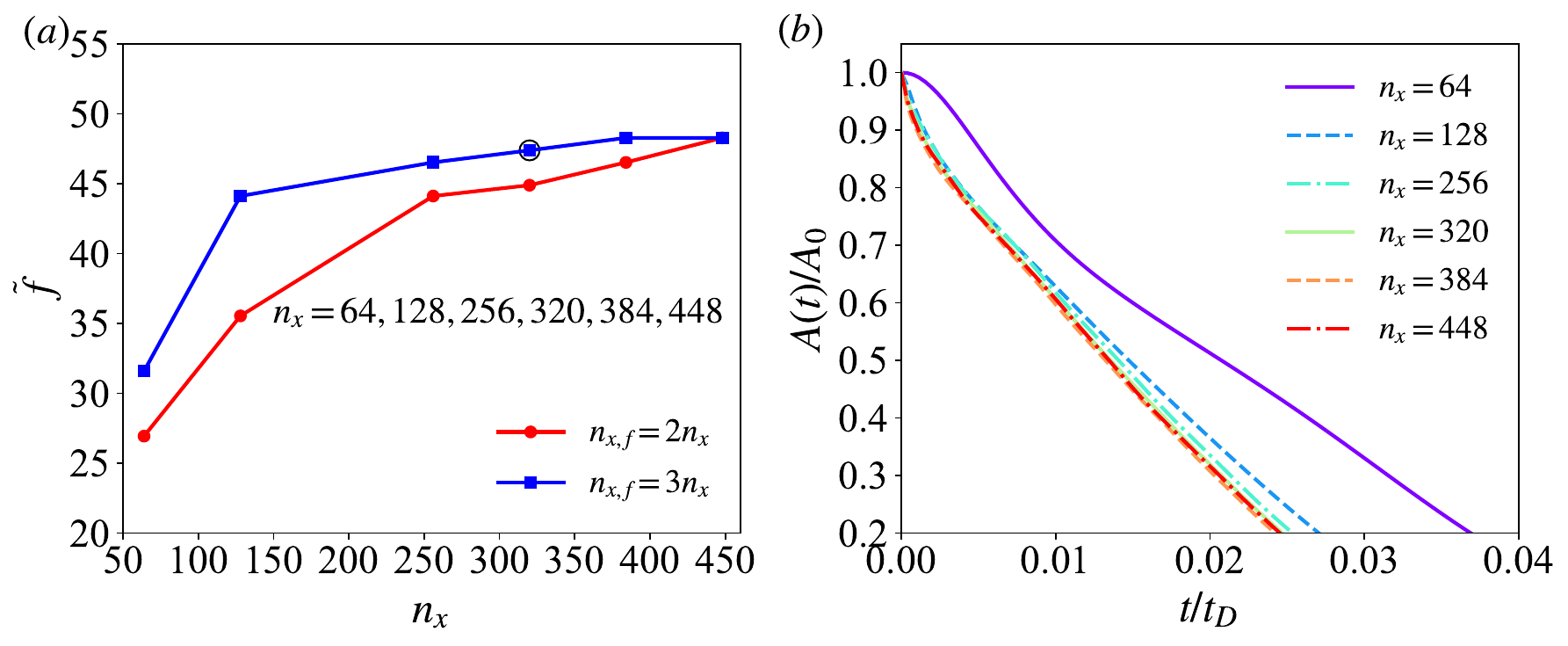}
    \caption{(\textit{a}) The normalised mean melt rate $\widetilde{f}$ as a function of the horizontal grid resolution $n_{x}$ for the velocity and temperature fields in the 2D case of $\text{Ra}_{T}=2.2\times 10^{5}$ and $\Delta S=\unit{0}{\gram\per\kilo\gram}$. The refined horizontal grid resolution $n_{x,f}$ is employed for the salinity and phase fields. Convergence is achieved as $n_{x}$ increases. In this case, we choose $n_{x}=320$ and $n_{x,f}=3n_{x}$ as shown by the black circle. (\textit{b}) The normalised area of ice $A(t)/A_{0}$ as a function of time $t/t_{D}$ in 2D case of $\text{Ra}_{T}=2.2\times 10^{5}$ and $\Delta S=\unit{0}{\gram\per\kilo\gram}$. The refined resolution $n_{x,f}$ is fixed to be $3n_{x}$. Here $A_{0}$ is the initial area of ice.}
\label{fig: app_a}
\end{figure}

Grid convergence tests were conducted for multiple cases to ensure that the same results are obtained. For brevity, only the grid-independence test for the 2D case with $\text{Ra}_T = 2.2\times 10^{5}$ and $\Delta S=\unit{0}{\gram\per\kilo\gram}$ is shown in figure \ref{fig: app_a}. Here $n_{x}$ denotes the horizontal grid resolution for the velocity and temperature fields, and $n_{x,f}$ represents the refined horizontal grid resolution for the salinity and phase fields. The results demonstrate convergence with increasing grid resolution. Therefore, we selected $n_x = 320$ and $n_{x,f} = 3n_x$ as the grid resolutions, as indicated by the black circle in figure \ref{fig: app_a}(\textit{a}).

\section{The influence of different Prandtl and Schmidt numbers}\label{appB}
\begin{figure}\centering
    \includegraphics[width=0.98\linewidth]{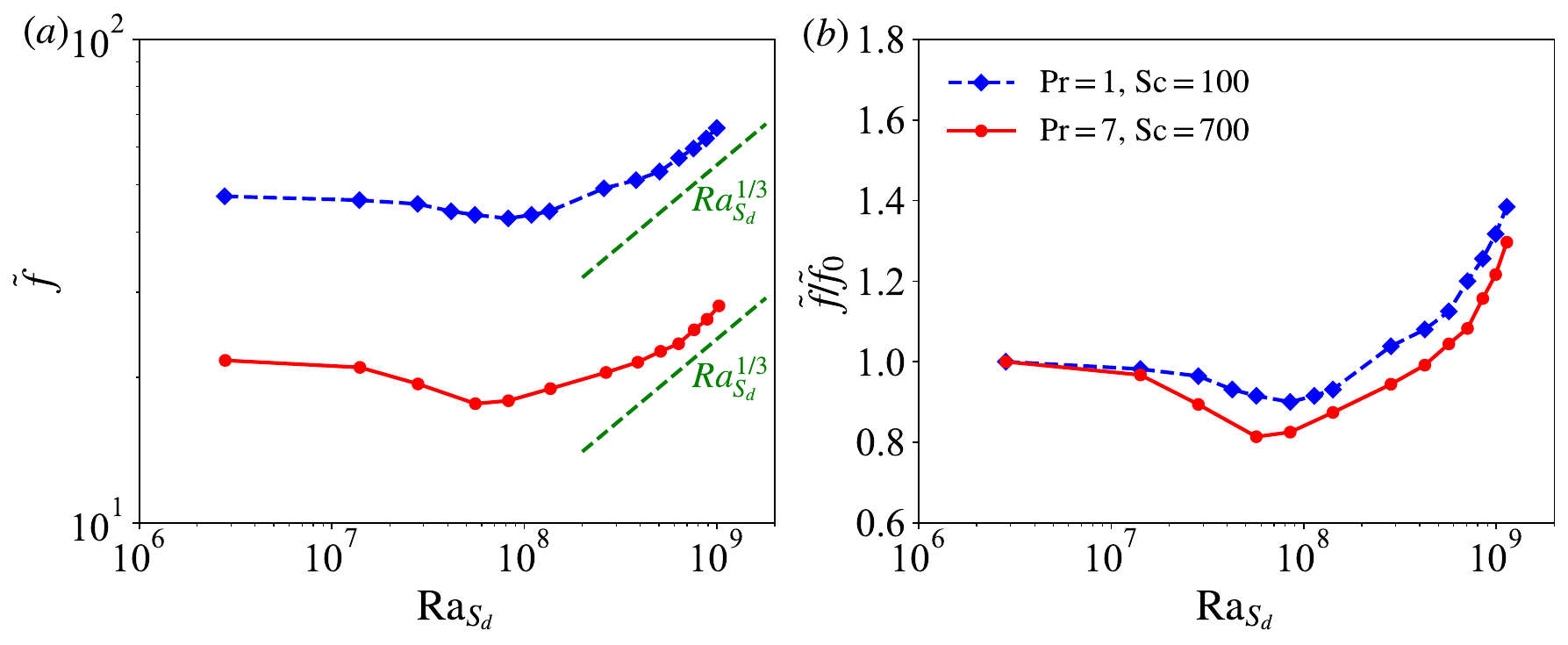}
\caption{(\textit{a}) The normalised mean melt rate $\widetilde{f}$ as a function of $\text{Ra}_{S_d}$ at $\text{Ra}_{T}=2.2\times 10^{5}$ for different Prandtl and Schmidt numbers. (\textit{b}) The relative mean melt rate $\widetilde{f}/\widetilde{f}_{0}$ as a function of $\text{Ra}_{S_d}$} at $\text{Ra}_{T}=2.2\times 10^{5}$ for different Prandtl and Schmidt numbers. Here $\widetilde{f}_{0}$ is the mean melt rate in freshwater ($\Delta S=0 \,\mathrm{g/kg}$).
\label{fig: app_b}
\end{figure}

The influence of different Prandtl and Schmidt numbers is examined in this Appendix. Figure \ref{fig: app_b} presents the normalised mean melt rate $\widetilde{f}$ and the relative mean melt rate $\widetilde{f}/\widetilde{f}_{0}$ as a function of $\text{Ra}_{S}$ at $\text{Ra}_{T}=2.2\times 10^{5}$ for different Prandtl and Schmidt numbers. Here, $\widetilde{f}_{0}$ represents the mean melt rate in freshwater ($\Delta S=\unit{0}{\gram\per\kilo\gram}$). It is noted that the grid resolutions required for high Prandtl and Schmidt numbers ($\text{Pr}=7$ and $\text{Sc}=700$) are significantly finer than those for lower values ($\text{Pr}=1$ and $\text{Sc}=100$). Specifically, for the 2D cases with $H=\unit{1.5}{\centi\meter}$, $\text{Ra}_T=2.2\times 10^{5}$, $\text{Pr}=7$ and $\text{Sc}=700$, a uniform mesh of $n_{x}\times n_{y}=256\times 512$ is employed for the velocity and temperature fields, and a uniform mesh for the salinity and phase fields increases from $n_{x,f}\times n_{y,f}=768\times 1536$ to $1536\times 3072$ as $\Delta S$ rises from $\unit{0}{\gram\per\kilo\gram}$ to $\unit{80}{\gram\per\kilo\gram}$. The results demonstrate that the mean melt rates $\widetilde{f}$ are significantly lower for high Prandtl and Schmidt numbers ($\text{Pr}=7$ and $\text{Sc}=700$) compared to those for low Prandtl and Schmidt numbers ($\text{Pr}=1$ and $\text{Sc}=100$). This reduction in mean melt rates $\widetilde{f}$ at higher Prandtl and Schmidt numbers is attributed to the smaller diffusivities of temperature and salinity under these conditions. Moreover, the relative mean melt rates  $\widetilde{f}/\widetilde{f}_{0}$ for high Prandtl and Schmidt numbers are slightly smaller than those for low Prandtl and Schmidt numbers in both the competing and salinity-driven regimes. Nevertheless, despite these reductions, the mean melt rates for high Prandtl and Schmidt numbers retain their non-monotonic dependence on ambient salinity and follow a $\text{Ra}_{S_d}^{1/3}$ scaling in the salinity-driven flow regime, in agreement with the trends observed at lower Prandtl and Schmidt numbers. Consequently, the use of lower Prandtl and Schmidt numbers in the simulations has a negligible impact on the qualitative findings of the present study.

\section{The instantaneous melt rates}\label{appC}

\begin{figure}\centering
    \includegraphics[width=0.98\linewidth]{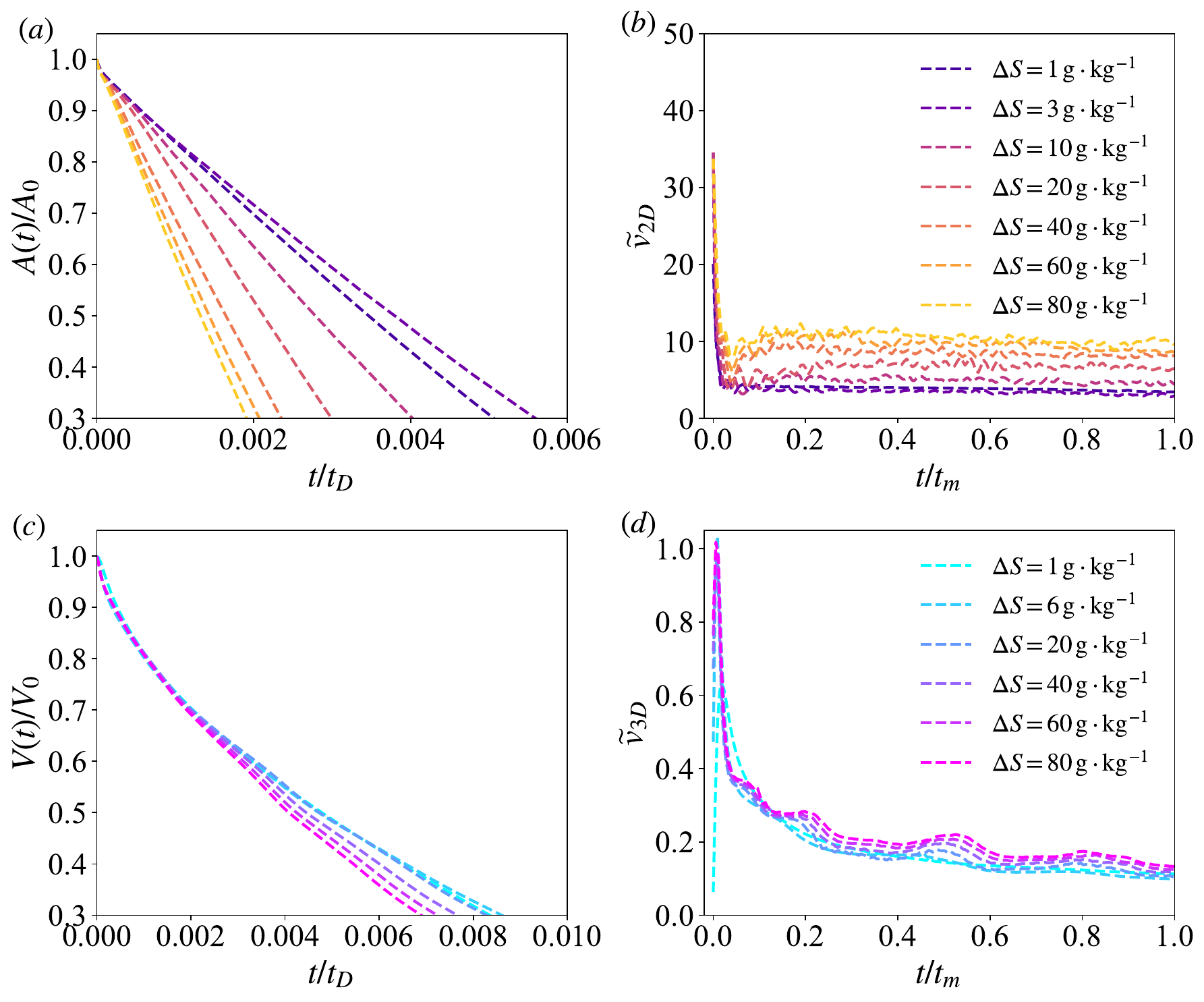}
\caption{(\textit{a}) The normalised area of ice $A(t)/A_{0}$ as a function of the normalised time $t/t_{D}$, and (\textit{b}) the instantaneous melt rate $\widetilde{v}_{2D}$ as a function of the normalised time $t/t_{m}$ for 2D cases with $\text{Ra}_{T}=1.1\times 10^{8}$ and varying ambient salinity $\Delta S$. (\textit{c}) The normalised volume of ice $V(t)/V_{0}$ as a function of the normalised time $t/t_{D}$, and (\textit{d}) the instantaneous melt rate $\widetilde{v}_{3D}$ as a function of the normalised time $t/t_{m}$ for 3D cases with $\text{Ra}_{T}=2.2\times 10^{5}$ and different ambient salinity $\Delta S$. Here, $A{0}$ and $V_{0}$ denote the initial ice area and volume in the 2D and 3D simulations, respectively, while $t_{m}$ represents the time required to melt $V_{m}=70\%$ of the initial ice area in 2D simulations and of the initial ice volume in 3D simulations.}
\label{fig: app_c}
\end{figure}

The instantaneous melt rates in both 2D and 3D simulations are investigated in this Appendix. Figure \ref{fig: app_c} presents the temporal evolution of the normalised ice area and volume, along with the corresponding instantaneous melt rates, $\widetilde{v}_{2D}$ and $\widetilde{v}_{3D}$, for the 2D and 3D cases, respectively. These melt rates are defined as $\widetilde{v}_{2D} = (t_D/A_0) dA/dt$ and $\widetilde{v}_{3D} = (t_D/V_0) dV/dt$. In 2D simulations, aside from an initially larger instantaneous melt rate due to the higher temperature gradients near the ice surface imposed by the initial conditions, $\widetilde{v}_{2D}$ remains approximately constant and attains a quasi-steady state within the range $0.2 \leq t/t_m \leq 1.0$. In contrast, in 3D simulations, the instantaneous melt rate $\widetilde{v}_{3D}$ slightly decreases over time as the ice melts within the same time interval.

\section{Vertical dependence of melt rates and boundary layer thicknesses}\label{appD}

\begin{figure}
    \centering
    \includegraphics[width=0.8\linewidth]{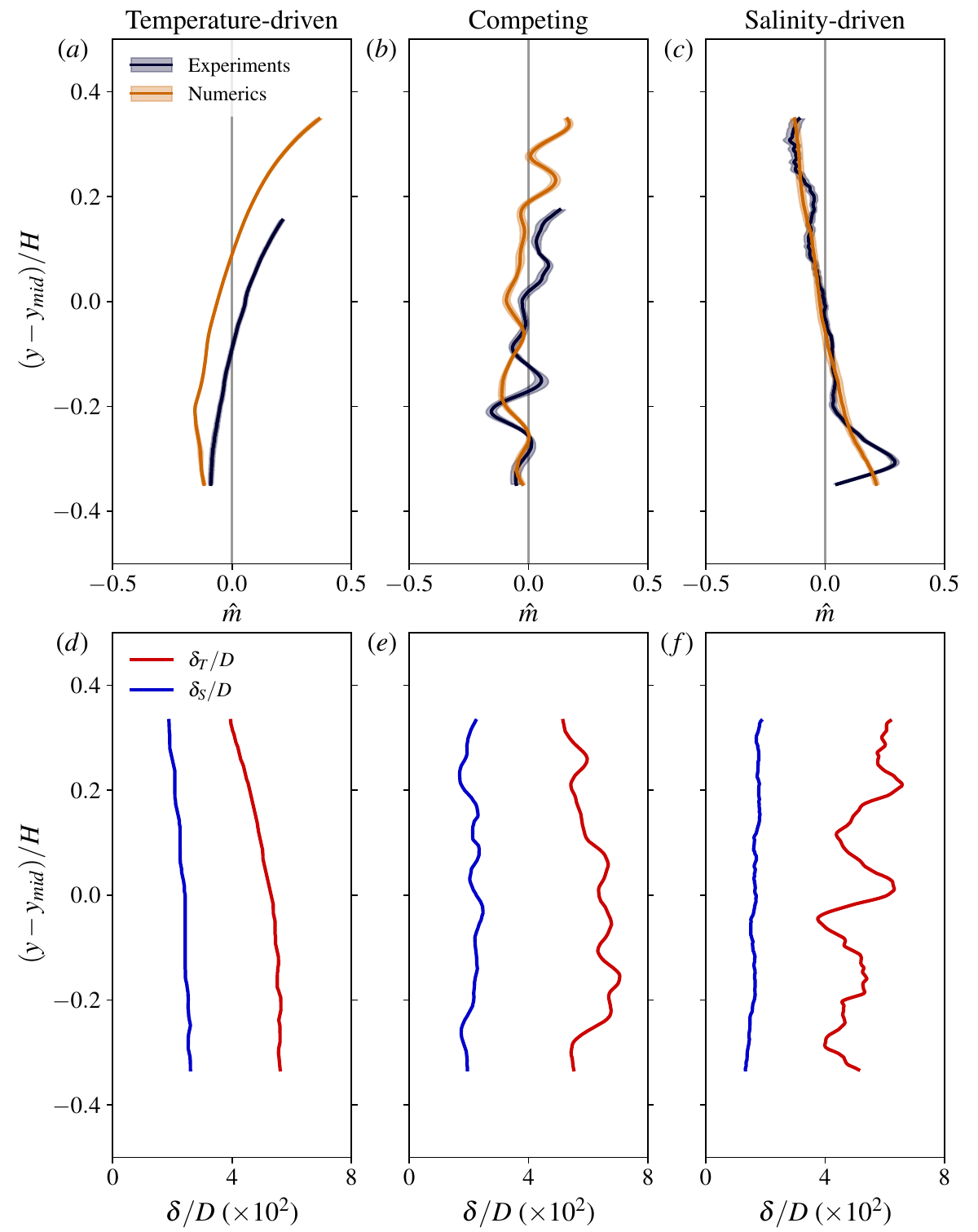}
    \caption{Time-averaged vertical profiles of (\textit{a}--\textit{c}) the normalised melt rate and (\textit{d}--\textit{f}) the thermal and saline boundary layer thicknesses for the (\textit{a},\textit{d}) temperature-driven, (\textit{b},\textit{e}) competing and (\textit{c},\textit{f}) salinity-driven flow regimes. The cases correspond to those shown in figures \ref{fig: fig6} and \ref{fig: fig7}. Only the middle part of the cylinder is shown to remove end effects at the top and bottom. }
    \label{fig: app_d1}
\end{figure}

The ice morphology for each of the flow regimes, as discussed in \ref{sec:flow_regimes}, is the result of a change in melt rate and thermal and saline boundary layer thicknesses along the cylinder. In figure \ref{fig: app_d1}, we show the vertical melt rate and boundary layer thickness profiles that further explain the observed morphology in figure \ref{fig: fig6}. Here the boundary layer thickness is computed as the \textit{e}-folding distance from the ice-water interface. 
In the temperature-driven regime (figure \ref{fig: app_d1}\textit{a},\textit{d}), the horizontal melt rate increases with the vertical coordinate for the largest part of the cylinder, leading to faster melting at the top compared to the bottom, as discussed in section \ref{sec:flow_regimes}. Both the thermal and saline boundary layer thicknesses increase with distance from the top, as a result of the downward flow of fresh and cold meltwater. 
For the competing flow regime (figure \ref{fig: app_d1}\textit{b},\textit{e}), the melt rate profiles show significant oscillations with height, due to the local variations in melt rate associated with scalloping, as discussed in section \ref{sec:morphology}. The boundary layers show similar oscillations, due to local accumulation of meltwater and entrainment of ambient water.
For the salinity-driven regime (figure \ref{fig: app_d1}\textit{c},\textit{f}), the melt rate generally decreases with the vertical coordinate, resulting from the salinity-driven upward flow of meltwater. An exception is the maximum melt rate around $y - y_\text{mid} \approx -0.3 H$, corresponding to the intrusion of ambient water due to the transition between temperature-driven and salinity-driven flow, as discussed in section \ref{sec:flow_regimes}. This maximum melt rate causes the dent observed in figure \ref{fig: fig6}(\textit{c}) to appear. Both boundary layers increase in thickness with distance from the bottom, indicating that the thermal boundary layer is controlled by the salinity-driven upward flow. While the saline boundary layer appears smooth, the thermal boundary layer shows oscillations that indicate a chaotic bulk flow. For a turbulent meltwater plume, it has been shown that both the melt rate and the boundary layer thickness become independent of the vertical coordinate at some height \citep{Wells2008, Gayen2016, Mondal2019}. However, the melt rate and boundary layer thickness profiles in figure \ref{fig: app_d1}(\textit{c}, \textit{f}) clearly depend on the vertical coordinate. Likely, the height of our cylinders is too small for this behavior to be observed. In numerical simulations of a dissolving vertical ice wall by \citet{Gayen2016}, the dissolution rate reached a maximum at approximately 4 cm from the bottom. With increasing distance from the bottom, the dissolution rate first decreased, before reaching a constant value at approximately 30 cm. This is consistent with our observations in figure \ref{fig: app_d1}(\textit{c}).

\FloatBarrier

\bibliographystyle{jfm}
\bibliography{jfm}



\end{document}